\documentclass[acmsmall]{acmart}

\AtBeginDocument{%
  \providecommand\BibTeX{{%
    \normalfont B\kern-0.5em{\scshape i\kern-0.25em b}\kern-0.8em\TeX}}}

\setcopyright{cc}
\setcctype{by}
\acmJournal{PACMHCI}
\acmYear{2025} \acmVolume{9} \acmNumber{5} \acmArticle{pn5758} \acmMonth{9} \acmPrice{}\acmDOI{10.1145/3743709}

\usepackage{booktabs}
\usepackage{todonotes}

\usepackage{subcaption} 
\usepackage{graphicx}
\usepackage{array}
\usepackage{rotating}
\usepackage{tabularx}
\usepackage{url}
\usepackage{xcolor}
\usepackage{booktabs}
\usepackage{verbatim}
\makeatletter
\g@addto@macro{\UrlBreaks}{\UrlOrds}
\makeatother
\usepackage{multirow} 
\usepackage{color} 
\usepackage{enumitem} 

\usepackage{longtable}
\usepackage{xspace}
\usepackage{tcolorbox}
\usepackage{hyperref}

\makeatletter
\renewenvironment{quote}
  {\list{}{\leftmargin3pt\rightmargin0pt}%
   \item\relax}
  {\endlist}
\makeatother

\newcommand{\chisq}{$\chi^2$}
\newcommand{\pminor}{\textit{p$<$}}
\newcommand{\F}[3]{$F({#1},{#2})={#3}$}

\newtcbox{\highlight}[1][magenta]{on line, arc=0pt,colback=#1!10!white,colframe=#1!50!black, before upper={\rule[-3pt]{0pt}{10pt}},boxrule=1pt, boxsep=0pt,left=4pt,right=3pt,top=2pt,bottom=1pt}

\newcommand{\notextrefrq}[1]{\hyperref[rq:#1]{\highlight{\textbf{\texttt{\textcolor{purple}{#1}}}}}}

\newcommand{\m}{\textit{M = }}

\newcommand{\sd}{\textit{SD = }}

\newcommand{\p}{\textit{p=}}

\usepackage{duckuments}
\usepackage{arydshln}

\begin{document}

\title[Exploring the Impact of Dark Patterns in MR Scenarios]{Mind Games! Exploring the Impact of Dark Patterns in Mixed Reality Scenarios}

\author{Luca-Maxim Meinhardt}
\email{luca.meinhardt@uni-ulm.de}
\orcid{0000-0002-9524-4926}
\affiliation{%
  \institution{Ulm University}
  \city{Ulm}
  \country{Germany}
}

\author{Simon Demharter}
\orcid{0000-0002-8768-2942}
\email{simon.demharter@uni-ulm.de}
\affiliation{%
  \institution{Ulm University}
  \city{Ulm}
  \country{Germany}
}

\author{Michael Rietzler}
\email{michael.rietzler@uni-ulm.de}
\orcid{0000-0003-2599-8308}
\affiliation{%
  \institution{Ulm University}
  \city{Ulm}
  \country{Germany}
}

\author{Mark Colley}
\email{m.colley@ucl.ac.uk}
\orcid{0000-0001-5207-5029}
\affiliation{%
  \institution{Ulm University}
  \city{Ulm}
  \country{Germany}
}
\affiliation{%
  \institution{UCL Interaction Centre}
  \city{London}
  \country{United Kingdom}
}

\author{Thomas Eßmeyer}
\email{mildner@uni-bremen.de}
\orcid{0000-0002-1712-0741}
\affiliation{%
  \institution{University of Bremen}
  \city{Bremen}
  \country{Germany}
}

\author{Enrico Rukzio}
\email{enrico.rukzio@uni-ulm.de}
\orcid{0000-0002-4213-2226}
\affiliation{%
  \institution{Ulm University}
  \city{Ulm}
  \country{Germany}
}

\renewcommand{\shortauthors}{Meinhardt et al.}

\begin{abstract}
Mixed Reality (MR) integrates virtual objects with the real world, offering potential but raising concerns about misuse through dark patterns. This study explored the effects of four dark patterns, adapted from prior research, and applied to MR across three targets: places, products, and people. In a two-factorial within-subject study with 74 participants, we analyzed 13 videos simulating MR experiences during a city walk. Results show that all dark patterns significantly reduced user comfort, increased reactance, and decreased the intention to use MR glasses, with the most disruptive effects linked to personal or monetary manipulation. Additionally, the dark patterns of \textit{Emotional and Sensory Manipulation} and \textit{Hiding Information} produced similar impacts on the user in MR, suggesting a re-evaluation of current classifications to go beyond deceptive design techniques. Our findings highlight the importance of developing ethical design guidelines and tools to detect and prevent dark patterns as immersive technologies continue to evolve.
\end{abstract}



\begin{teaserfigure}
\centering
 \includegraphics[width=0.99\textwidth]{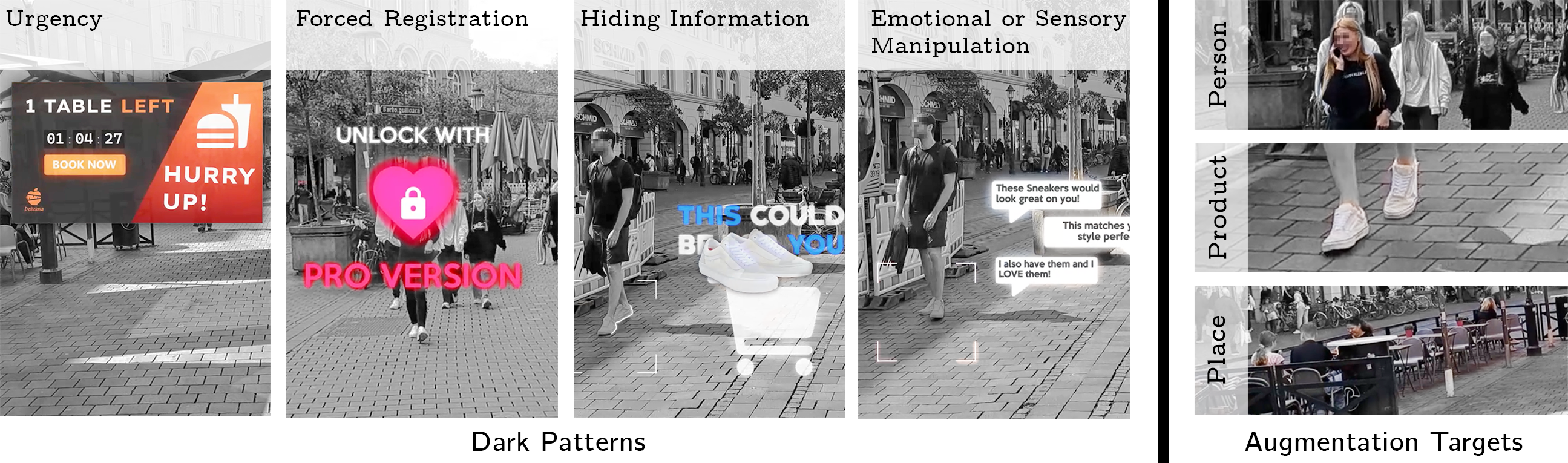}
  \caption{Overview of the four dark patterns adopted to MR and the target that they were applied to.}
  \Description{This figure illustrates four dark patterns in MR and their corresponding augmentation targets (a person, a product (sneakers), and a place). On the left, the dark patterns section highlights the following deceptive design techniques: (1) Urgency – an MR advertisement uses a countdown timer and phrases like "1 TABLE LEFT" and "Hurry up!" to create pressure for immediate action to book a table at a restaurant; (2) Forced Registration – an MR heart hiding a person in a public scene requires users to upgrade or register for access, displaying "UNLOCK WITH PRO VERSION" along with a lock icon; (3) Hiding Information – the sneakers are sneaked into the users shopping basket without prior consent; and (4) Emotional or Sensory Manipulation – augmented messages, such as "These sneakers would look great on you!" or "I also have them, and I LOVE them!" are used to create emotional connections and social pressure.}
  \label{fig:teaser}
\end{teaserfigure}
\maketitle

\section{Introduction}\label{sec:intro}

Mixed Reality (MR) blends the virtual and physical worlds, integrating aspects of both augmented and virtual reality as defined by \citet{milgram_augmented_1995}. 
As this technology advances, MR drives new applications in fields like education~\cite{3_towards_2022, Jin_2022} and entertainment~\cite{min_vpmodel_2023}. However, the current attention on MR has also intensified concerns about the potential misuse of this technology, envisioning a dystopian near future~\cite{eghtebas_co-speculating_2023,johnstone_when_2022, wang_dark_2023}. For instance, the ubiquity of MR interfaces holds the potential for psychological manipulation~\cite{bonnail_memory_2023}, misinformation~\cite{rixen_may_2023}, and adverse advertisements~\cite{mhaidli_identifying_2021}. Similar concerns were already raised a decade ago by \citet{greenberg_dark_2014}, whose speculative views on proxemic interactions highlighted the potentially deceptive aspects of ubiquitous technology, such as interactive digital advertising in urban areas. 

With the ongoing development of MR technology, exemplified by recent products like the Apple Vision Pro~\cite{vision_proApple} or Meta Orion~\cite{metaOrion}, the pervasiveness of this technology increases, and so does the responsibility of Human-Computer Interaction (HCI) practitioners to design technologies that align with ethical values~\cite{doorn_value_2013} and support moral reflection~\cite{verbeek2015toward}. However, as the Collingridge-Dilemma~\cite{collingridge_social_1982} illustrates, it is challenging to regulate new technology as the societal impact of technology is impossible to predict before it becomes widespread. Yet, the historical evolution of applications and user interfaces (UIs) revealed a rising prevalence of deceptive design practices, which have a detrimental impact on user experiences across various digital domains such as entertainment~\cite{kai_internal, Chiossi.2023}, social media~\cite{Mildner.2023, schaffner_understanding_2022, mongeroffarello2022towards, meinhardt_balancing_2023} and e-commerce~\cite{mathur_dark_2019}.
These deceptive design techniques, also known as \textit{dark patterns}\footnote{We use the term \textit{dark patterns} to align with previous research and legal definitions, despite alternatives like \textit{deceptive design} or \textit{manipulative design}. Although the \href{https://www.acm.org/diversity-inclusion/words-matter}{ACM Diversity and Inclusion Council} considers it potentially problematic, \textit{dark patterns} remains a well-established term to describe a wide range of hidden, deceptive, and coercive tactics that reduce users' agency and steer their behavior.}~\cite{gray_dark_2021, Gray.2018, gunawan_comparative_2021, mathur_dark_2019, Gray.2024}, are described as "\textit{[...] tricks used in websites and apps that make you do things that you did not mean to, like buying or signing up for something}"~\cite{brignull-2023}. 
Since the introduction of the term \textit{dark patterns} by \citet{brignull-2023} in 2010, several papers~\cite{Gray.2018, mathur_dark_2019, Gray.2024, bosch_tales_2016} have classified these patterns into groups such as \textit{Forced Action}, \textit{Obstruction}, \textit{Sneaking}, \textit{Interface Interference}, and \textit{Social Engineering}.
Just as dark patterns are used in mobile and web applications~\cite{brignull-2023}, the growing body of research~\citet{krauss_what_2024,tseng_dark_2022, hadan_deceived_2024} on dark patterns in MR suggests that the embedding of these patterns will similarly exploit this new technology.
For instance, \citet{krauss_what_2024} explored dark patterns through co-design workshops, finding that most of them are new representations of known patterns. However, they state that due to the high immersion of this new technology, these new representations are likely to create higher deception compared to established UIs~\cite{krauss_what_2024}. The potential for such deception becomes even more substantial in the context of the Metaverse, where dark patterns could be used to exert pervasive influence over consumers~\cite{kim_advertising_2021, buhalis_metaverse_2023}. MR allows digital content to be overlaid on various real-world targets. For instance, \citet{mhaidli_identifying_2021} demonstrated how MR can augment specific objects, such as a fashion brand's logo on a t-shirt~\cite{bonner_when_2023}. The idea of using MR in advertising contexts is not new, as work explores how underlying technologies can be used to maximize the engagement of potential customers~\cite{scholz_augmented_2016}. This capability to individually target different aspects of the physical world to augment customized content for any bystander raises critical ethical concerns about an audience's agency not becoming a targeted subject~\cite{greenberg_dark_2014}. These concerns might further vary depending on whether the augmentation is applied to different targets, such as a \textit{place}, \textit{product}, or \textit{person}. In this work, we opted for these three domains to reflect frequent commercial targets~\cite{singh_impact_2024, bonner_when_2023}. Further, each example represents a distinct category: physical environments, individual products, and human subjects.

Despite growing research on dark patterns in MR~\cite{krauss_what_2024, wang_dark_2023, mhaidli_identifying_2021, hadan_deceived_2024}, there is still a lack of understanding about how these patterns impact users in everyday scenarios. Most studies rely on qualitative methods l  ike speculative co-design workshops~\cite{krauss_what_2024, todhriimmersion} aiming for hypothetical scenario creation, not focusing on the impact on users through these scenarios.
To address this gap, we transition from theoretical and qualitative methods to a scenario-based quantitative study, examining how dark patterns influence users in simulated MR environments. MR's ability to seamlessly integrate virtual objects into the real world creates highly immersive experiences, making it an ideal platform for studying the nuanced effects of manipulative design techniques. 
Thus, this study is guided by the research question (RQ):
\begin{quote}
      \textbf{What impact do dark patterns in Mixed Reality have on users when augmented to different targets (place, person, product)?}
\end{quote}

\noindent We conducted a two-factorial online user study with N=74 participants, assessing 13 videos showing a city walk through a pedestrian zone from the perspective of an individual wearing MR glasses. These videos were structured into a 4 $\times$ 3 + 1 baseline factorial design. The factors included four dark patterns (\textit{Emotional or Sensory Manipulation}, \textit{Forced Registration}, \textit{Hiding Information}, and \textit{Urgency}), corresponding to the meso-level patterns identified by \citet{Gray.2024}, who noted that the meso-level is context agnostic and can thus be applied to application types including MR. Following their recommendations, we interpreted the dark patterns based on "\textit{the specific context of use or application type}"~\cite[p. 8]{Gray.2024}. Consequently, the dark patterns were applied to three different targets as contexts: a place (a restaurant in the pedestrian zone), a product (shoes of a pedestrian), and a person--which we derived from related work~\cite{Rixen2, singh_impact_2024, bonner_when_2023}. Additionally, a baseline condition was presented without any dark patterns. For each video, we assessed the participants' subjective ratings, such as reactance and level of comfort.

Our results revealed that all dark patterns in MR significantly lowered ratings for comfort~\cite{rixen_exploring_2021} and intention to use~\cite{venkatesh_theoretical_2000} while increasing users' reactance~\cite{Ehrenbrink.2020} and system darkness~\cite{van_nimwegen_shedding_2022} compared to the baseline.
The dark pattern of \textit{Forced Registration} by obscuring the user's face (\textit{Person}) and hiding information to sneak products into the shopping cart (\textit{Product}) was particularly disruptive, eliciting strong negative reactions in the open feedback. Hence, we assume that personal interference and monetary intentions were perceived as particularly disruptive. 
Further, our findings indicate that in MR, the impact of the dark patterns \textit{Emotional and Sensory Manipulation} and \textit{Hiding Information} on user reactance and comfort is very similar concerning the augmentation targets used in the study, suggesting that current classifications of dark patterns, which focus on design techniques, may also need to consider their actual effects on users. 



\medskip

\noindent\highlight[purple]{Contribution Statement~\cite{Wobbrock.2016}}

    \medskip
    \noindent\textbf{Empirical study that tells us about people.} We conducted a within-subject online study with N=74 participants, using quantitative measures to assess the impact of four dark patterns applied to three augmentation targets. All four dark patterns significantly reduced comfort, increased reactance and system darkness, and decreased intention to use MR glasses. \textit{Forced Registration} and \textit{Hiding Information} were the most disruptive, highlighting the impact of personal interference and monetary manipulation. The similar effects of \textit{Emotional or Sensory Manipulation} and \textit{Hiding Information} suggest reconsidering dark pattern classifications based on user impact.

\section{Related Work}
We adopt the definition of Mixed Reality (MR) proposed by \citet{milgram_augmented_1995}, which describes MR as an environment "\textit{[...] in which real world and virtual world objects are presented together within a single display, that is, anywhere between the extrema of the [Reality-Virtuality] continuum.}"\cite[p. 2]{milgram_augmented_1995}. While this definition is partly consistent with more recent interpretations~\cite{speicher_MR_definition_2019}, other researchers advocate for the term xReality (XR) as an umbrella term for all new forms of reality (including AR, VR, and MR), with \textit{X} serving as a placeholder for any reality format~\cite{rauschnabel_what_2022}. In this section, we provide an overview of related work on dark patterns in general and in XR. We refer to the terminology used by each paper, noting whether they discuss AR/VR specifically or address MR/XR in a broader sense.


\subsection{Dark Patterns}
Dark patterns were first introduced by \citet{brignull-2023} in 2010. They refer to interface designs within websites or mobile applications that manipulate users by steering, or deceiving users into making unintended and potentially harmful decisions, often leading to increased purchases or disclosure of information~\cite{mathur_what_2021, mathur_dark_2019}. 
In light of these issues, research resulted in the development of multiple classifications of these patterns~\cite{Gray.2018, mathur_dark_2019, bosch_tales_2016, luguri_shining_2021}. 
\citet{Gray.2024} integrated these classifications into an ontology, organizing dark patterns into five groups (\textit{Obstruction}, \textit{Sneaking}, \textit{Interface Interference}, \textit{Forced Action}, and \textit{Social Engineering}). They further establish a three-level hierarchy for each group, which distinguishes between high-level (the general strategy of manipulation), meso-level (the context-agnostic use of the application type) and low-level (the specific mean of execution).
The severe harm imposed by deceptive design is mirrored by legislative bodies who have begun to regulate the use of dark patterns on consumer websites, such as the European Commission through the Digital Services Act~\cite{DSA} and the Federal Trade Commission~\cite{FTC}. However, despite measures to prohibit dark patterns~\cite[§67]{DSA}, these practices remain widespread, as, for example, many websites still fail to comply with basic EU consent banner requirements~\cite{nouwens_dark_2020, gray_dark_2021}.


However, dark patterns do not only occur on consumer websites but also in social media applications~\cite{Mildner.2023} designed to limit the users' agency and to control their decision-making. For instance, \citet{mongeroffarello2022towards} specified infinite scrolling as a dark pattern in social media platforms that aims to engage users to stay longer in the social media application by constantly loading new content while scrolling. 
Other emerging technologies also contain dark patterns. Mobile devices containing conversational UIs (CUIs), for instance, have been argued to possibly misalign expectations between users and systems by anthropomorphic features overstating the system's capabilities~\cite{Mildner.2024} that can lead to deceptive interactions with CUIs~\cite{owens_exploring_2022}. Similar concerns were voiced in the context of Internet-of-Things~\cite{kowalczyk_understanding_2023}. 
Thus, these works highlight the importance of understanding how dark patterns manifest beyond traditional UIs.

Relevant to understanding possible dark patterns in XR by exploring proxemic interactions, \citet{greenberg_dark_2014} first envisioned and discussed dark patterns in ubiquitous commercial interfaces such as interactive billboards. This work presents examples of how public digital interfaces could exploit people's proximity depending on context-specific situations. For instance, systems could collect people's data without them noticing or displaying targeted advertisements publicly, raising serious privacy concerns. \citet{eghtebas_co-speculating_2023} extended these ideas to AR, showing how AR can intensify privacy invasions, and psychological harm by integrating virtual elements into physical spaces.
Thus, they advocate for proactive measures in designing and regulating AR technologies to address the identified risks. Further, they highlight that AR could amplify existing social issues or create new ones without careful consideration and intervention, leading to a dystopian future. 
Given the potential dark patterns inside future XR applications, the next section will examine research investigating possible manipulations in XR, which is an overarching term including AR, VR, and MR~\cite{rauschnabel_what_2022}.

%

\subsection{Dark Patterns and Manipulations in xReality}
Research has been conducted on manipulating perceptions in XR to enhance the immersive experience and extend the interaction. For instance, weight manipulation techniques create the illusion of lifting heavy objects in VR~\cite{rietzler_breaking_2018, stellmacher_continuous_2023}. Additionally, redirecting the user's walking path can manipulate them to walk in a small circle while perceiving themselves as walking straight~\cite{rietzler_rethinking_2018}. 

While these manipulations aim at positive application scenarios (such as reducing the users' operational space or increasing immersion), \citet{tseng_dark_2022} warns that such techniques in VR "\textit{[...] have the potential to be misused to provoke physical harm in the future [...]}"~\cite[p. 12]{tseng_dark_2022}. They illustrate how methods like redirected walking could be subverted to guide users toward physically hazardous locations, a category they named \emph{Puppetry Attacks}. Additionally, they define \emph{Mismatching Attacks}, in which an intended mismatch between virtual and real environments (e.g. a missing stair step) leads users to trip or fall.
Shifting towards AR, \citet{wang_dark_2023} explored manipulation techniques used to influence user interactions with virtual menus in AR. Their study revealed that visual manipulations, such as positioning the virtual menus behind objects (e.g., a street light), significantly impacted participants' decisions, leading more of them to agree to share personal data when these interferences were present. 
Further, potential misinformation could pose additional threats for AR users, as participants in a user study by \citet{rixen_may_2023} tended to trust the information presented on an AR device over verbally communicated information by the individual. Visual misinformation also caused concerns, as shown in a study where participants were shown multiple AR filters~\cite{bonner_when_2023}. In this study, participants were worried about filters that changed a person's appearance, such as digital blackface or hiding pregnancy. While these visual and psychological manipulations can also be used for positive purposes, such as supporting helping behavior in complex social situations~\cite{rixen_exploring_2023} or assisting XR users in finding common conversation topics~\cite{nguyen_known_2015}, \citet{mhaidli_identifying_2021} demonstrated that XR can facilitate manipulative advertising techniques. By creating speculative scenarios, they illustrated how XR advertisements could exploit users, such as by inserting fake relatives into VR chats to give financial advice or by digitally augmenting a sports brand's logo onto every t-shirt recognized by AR glasses. From these hypothetical scenarios, they identified five ways XR advertising can be manipulative: misleading experience marketing, inducing artificial emotions in consumers, targeting consumers when they are vulnerable, emotional manipulation, and distorting reality~\cite{mhaidli_identifying_2021}. While their work focuses on manipulative advertising in XR and does not explicitly mention dark patterns, the described manipulations closely align with Brignull's definition of dark patterns~\cite{brignull-2023}.


In fact, recent studies explored how XR technology could intensify dark patterns as the line between reality and virtuality increasingly blurs. For instance, \citet{todhriimmersion} analyzed 80 XR applications for dark patterns and identified that their usage impacts users' decision-making. Further, a speculative design workshop~\cite{krauss_what_2024} revealed that the spatial placement of virtual objects could influence user behavior, e.g., creating virtual barriers can cause discrepancies between the real world and virtual objects, leading to confusion and potential manipulation. According to \citet{hadan_deceived_2024}, this manipulation extends to changes in users' beliefs and morals, as users have false confidence in the objectivity of XR~\cite{rixen_may_2023, hadan_deceived_2024}. In addition, they point to privacy and security concerns arising from XR's extensive data collection, which enables more severe personalized manipulations than their web or mobile counterparts~\cite{hadan_deceived_2024}. 
Nevertheless, \citet{hadan_deceived_2024} noted that most "\textit{[...] studies [of dark patterns in XR] were based on hypothetical scenarios of potential future problems}"~\cite[p. 18]{hadan_deceived_2024}, leaving a gap in our understanding of their actual impact on users. To address this limitation, this paper shifts the focus from qualitative, speculative approaches to a more quantitative exploration. By utilizing concrete visualizations of dark patterns in an everyday scenario, this work explores the impact of dark patterns on MR users. Thus, we conducted a video-based online study, described in the next section.



\section{User Study}
To evaluate the impact of dark patterns in MR based on their augmentation targets, we conducted a within-subject online video study (N=74) with a 4 $\times$ 3 + 1 baseline factorial design. 
Our first factor (\textit{Dark Pattern}) included four levels of different dark patterns: (1) \textit{Emotional or Sensory Manipulation}, (2) \textit{Forced Registration}, (3) \textit{Hiding Information}, and (4) \textit{Urgency}. These dark patterns were derived from the meso-level dark patterns described by \citet{Gray.2024}, who identified five high-level dark patterns and further divided them into concrete meso- and low-level patterns. We included these meso-level patterns in our user study because they "\textit{[...] may be interpreted in a contextually-appropriate way based on the specific context of use or application type}"~\cite[p. 8]{Gray.2024}. This allowed us to adapt the dark patterns to the specific context or target they were augmented onto, as defined as our sound factor in the user study.
This second factor (\textit{Augmentation Targets}) focused on the augmentation targets, with three levels: place, product, and person, inspired by \citet{rixen_exploring_2021}. We used these targets to explore how different contexts of augmentation might influence user reactions in MR scenarios, providing a range of targets that vary in their association with personal identity and privacy. Additionally, we included a baseline condition without any dark patterns for comparison. According to \citet{mathur_what_2021}, "\textit{[..] the appropriate baseline may be derived from the space of designs that are similar to the concerning UI but that do not contain the possible dark pattern.}"~\cite[p. 15]{mathur_what_2021}. Hence, in this baseline condition, participants were still exposed to MR elements such as navigation aids, weather forecasts, and contextual information about the environment, but none of the dark patterns were applied. This approach for the baseline visualizations was taken from \citet{rixen_exploring_2023}.

We performed an a-priori power analysis to determine the minimum sample size required to detect a medium effect size of 0.3~\cite{cohen_power_1992}. Using the \texttt{pwr} package version 1.3 in R~\cite{pwr_r_package}, we calculated that with a desired significant level of 0.05 and a power of 0.8, a sample size of 58 participants would be appropriate for our study design. 

\subsection{Materials}

\begin{figure*}[ht!]
    \centering
    \small
    \includegraphics[width=0.999\linewidth]{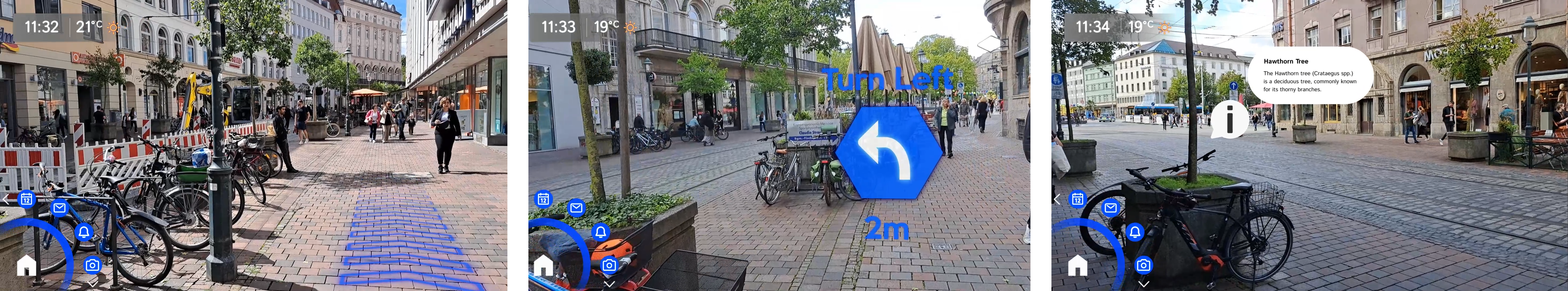}
    \caption{Baseline augmentations included navigation aids, weather forecasts, and contextual information about the environment. These augmentations were present in each condition.}\label{fig:study_baseline}
    \Description{This figure displays three baseline augmentations highlighted in turquoise. On the left, the bottom corner features a menu with four items, including a calendar and photo icon. In the pedestrian zone, an augmented trajectory is shown along the path. In the center, a navigation aid appears above the street, displaying an icon with the text "Turn Left 2m" and an arrow pointing left. On the right, additional environmental information is augmented next to a tree, providing details about the hawthorn tree in a small text box.}
\end{figure*}

We recorded a video of a city walk through the pedestrian zone in \anon{Augsburg, Germany}. Using this footage as the foundation, we created 13 two-minute videos with After Effects. These videos were designed to simulate what MR technology could look like, adapting the original footage to create an MR experience. Presenting the MR scenario via video format made the study accessible to a broader, more diverse participant pool~\cite{spielHowBetterGender2019, Peck2020Mind}, ensuring that individuals without specialized MR or VR headsets could still fully participate.


All videos featured augmentations like navigation aids, weather forecasts, and contextual information about the environment, following suggestions from \citet{rixen_exploring_2023} (see \autoref{fig:study_baseline}). In the baseline condition, these were the only augmentations visible, while in the other conditions, the videos additionally included augmented dark patterns.

\subsubsection{Dark Patterns} \label{sec:dark_patterns}
The videos included four meso-level dark patterns aligned with \citet{Gray.2024}. The following list describes which high-level dark pattern corresponds to each meso-level pattern, along with a brief description of each dark pattern according to \citet{Gray.2024}. While other classifications of dark patterns exist~\cite{Gray.2018, mathur_dark_2019, bosch_tales_2016, luguri_shining_2021}, the ontology by \citet{Gray.2024} is the most recent and comprehensive by considering over 200 types of dark patterns in a transdisciplinary context, drawing form five academic publications as well as five regulatory reports. These include but are not limited to, research by \citet{mathur_dark_2019}, \citet{luguri_shining_2021}, or \citet{Gray.2018}, and also reports by the FTC~\cite{FTC2022} and EDPB~\cite{edpb_guidelines_2022}, highlighting the need to protect end-users from dark pattern harms. Following a qualitative approach, the authors identified 64 unique strategies between these works, organizing them within a hierarchical structure spanning context-agnostic high- and meso-level dark patterns as well as context-specific low-level instances. However, we refrained from using the dark pattern of \textit{Obstruction}, as it "\textit{impedes a user's task flow}"~\cite[p. 17]{Gray.2024}. Hence, the meso-level dark patterns linked to the high-level strategy \textit{Obstruction} typically require a longer user journey to manifest (e.g., Roach Motel: making it difficult or impossible to delete a user account), whereas the other four dark patterns can be demonstrated within single UIs.

\begin{enumerate}
    \item \textbf{Emotional or Sensory Manipulation} (High-level category: \textit{Interface Interference})
    \begin{itemize}
        \item Alters language, style, color, or other design elements to evoke emotions or manipulate the senses, persuading the user to take a specific action.
    \end{itemize}

    \item \textbf{Forced Registration} (High-level category: \textit{Forced Action})
    \begin{itemize}
        \item Steers users into believing that registration is required to proceed, often making it difficult to access content or features without signing up or taking additional actions.
    \end{itemize}

    \item \textbf{Hiding Information} (High-level category: \textit{Sneaking})
    \begin{itemize}
        \item Conceals or delays important details until later in the user journey, potentially influencing the user's decision. For example, adding unwanted items to a user’s shopping cart without their consent.
    \end{itemize}

    \item \textbf{Urgency} (High-level category: \textit{Social Engineering})
    \begin{itemize}
        \item Creates a sense of urgency by indicating that a deal or discount will expire soon, typically through the use of a countdown clock or timer.
    \end{itemize}
\end{enumerate}


\begin{table*}[ht!]
\centering
\caption{Overview of the four dark patterns adopted to MR for this user study aligned with \citet{Gray.2024}. The dark patterns were augmented onto different targets (place, product, person)}
\Description{This table presents the four dark patterns and the three augmentation targets used in the video study. Each cell includes an image representing a dark pattern, with a description provided below the image.}
\label{tab:dark_patterns}
\scriptsize

\begin{tabular}[ht!]{|>{\centering\arraybackslash}p{1cm}|>{\centering\arraybackslash}p{3.7cm}>{\centering\arraybackslash}p{3.7cm}>{\centering\arraybackslash}p{3.7cm}}

\hline
\parbox[c][1cm][c]{1.2cm}{meso-lvl. \newline \textbf{dark \newline pattern}} & 
\parbox[c][1cm][c]{1.2cm}{\centering \textbf{Place}} & 
\parbox[c][1cm][c]{1.2cm}{\centering \textbf{Product}} & 
\parbox[c][1cm][c]{1.2cm}{\centering \textbf{Person}} \\
\hline
\addlinespace[2pt]
\rotatebox[origin=c]{90}{\textbf{Emotional Manipulation}} &
\parbox[c][3.1cm][t]{\linewidth}{%
\includegraphics[width=\linewidth]{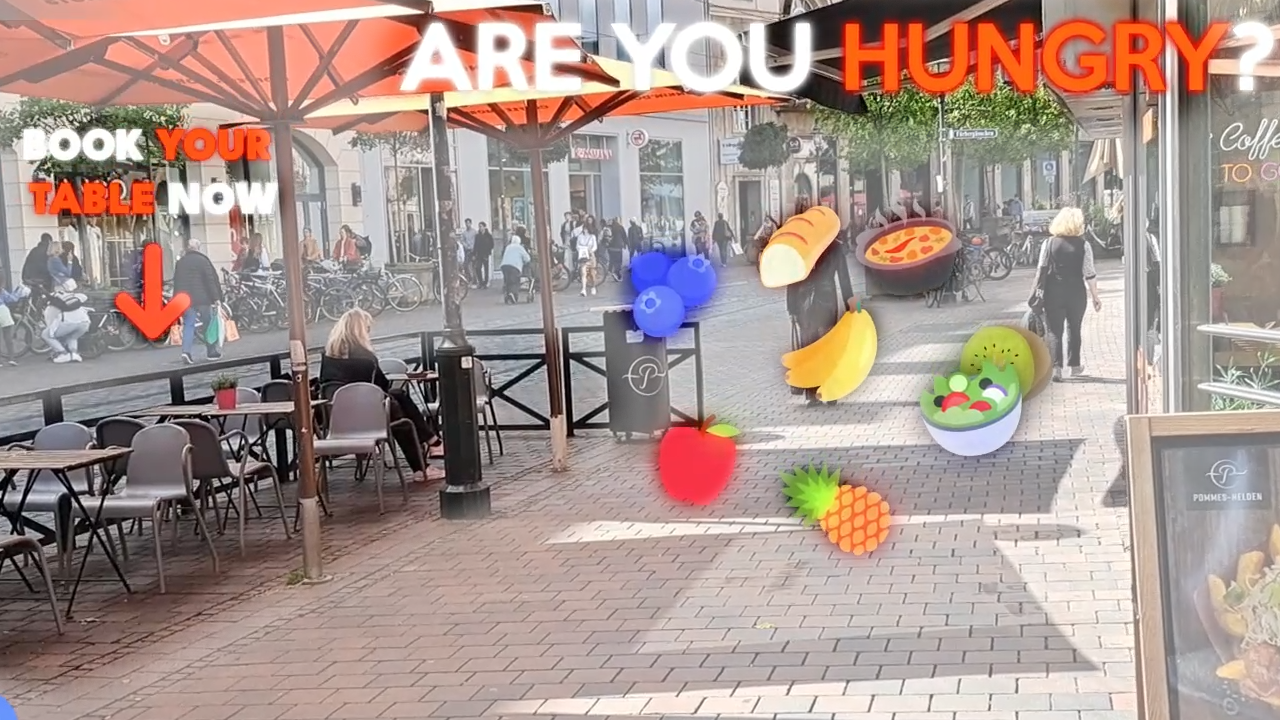} \newline 
Augmentation of food in front of the restaurant with the headline "Are you hungry?" induces emotional and sensory stimuli to sit and dine.} & 
\parbox[c][3.1cm][t]{\linewidth}{%
\includegraphics[width=\linewidth]{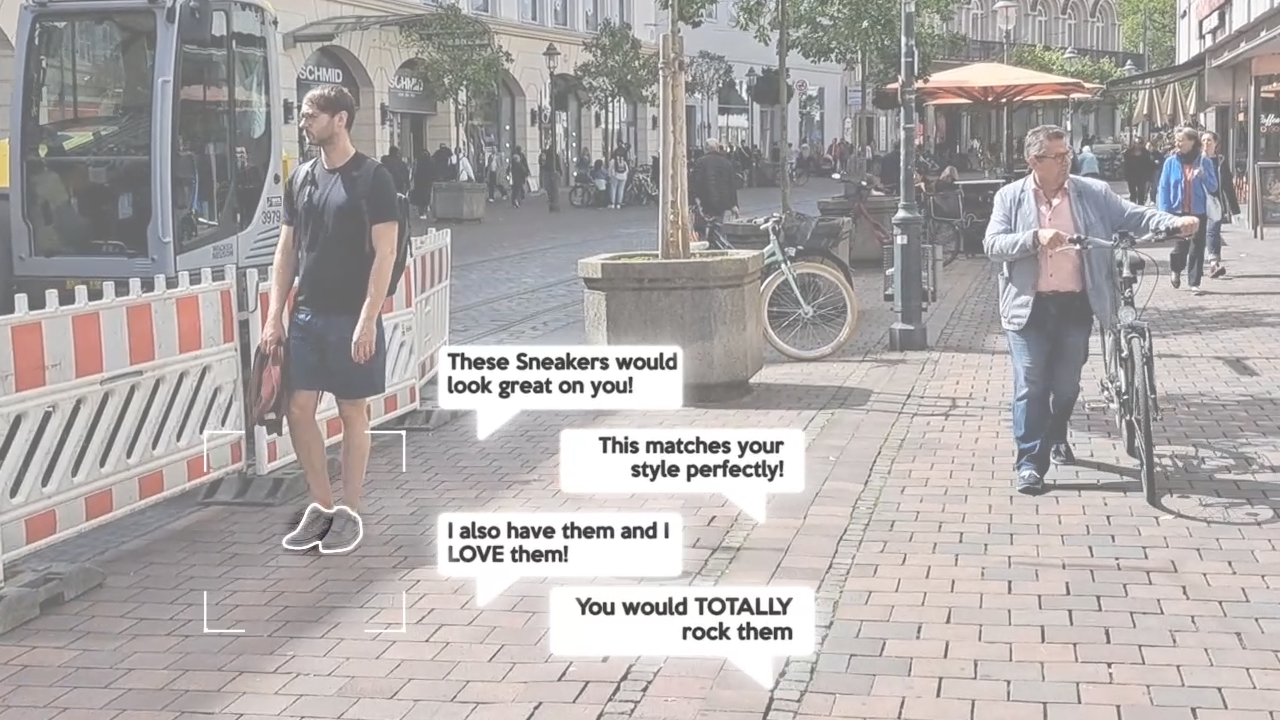} \newline 
The highlighted shoes were augmented with the text: "These shoes would look great on you!". This was intended to complement and persuade the viewer to buy the product.} & 
\parbox[c][3.1cm][t]{\linewidth}{%
\includegraphics[width=\linewidth]{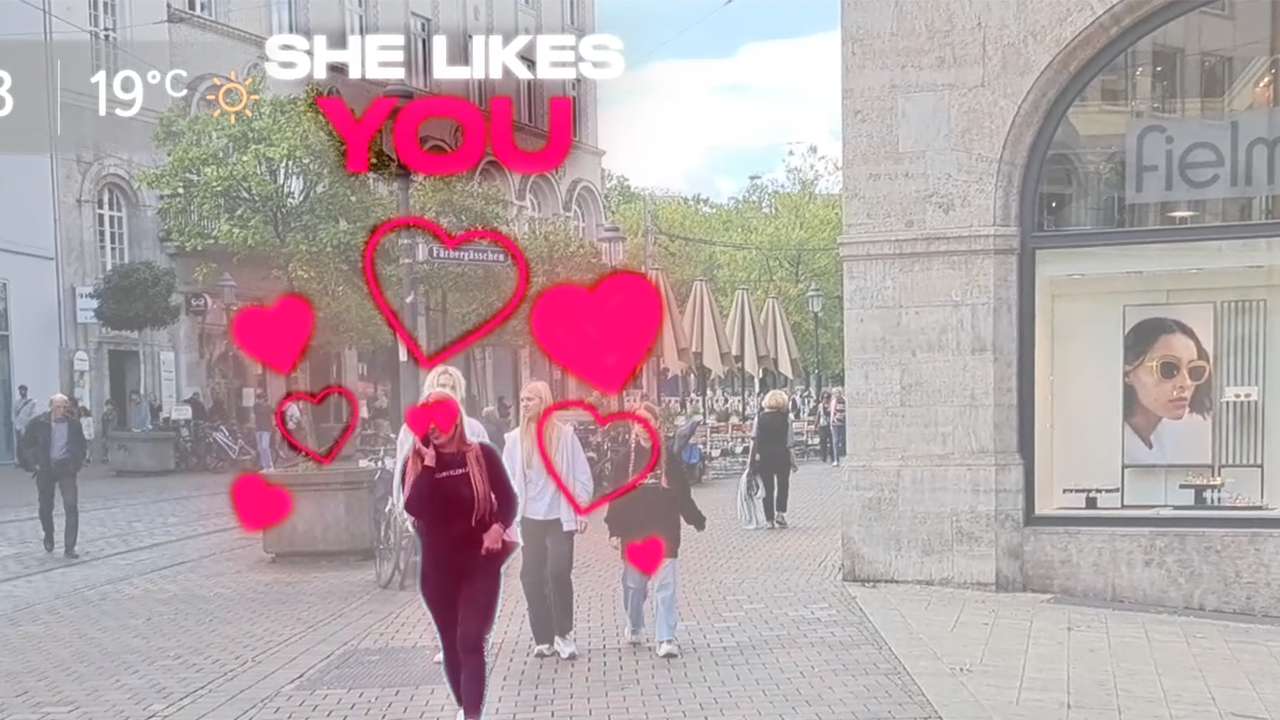} \newline 
The person was surrounded by animated hearts and the text: "She likes you" to create emotional manipulation.} \\[15pt]

\rotatebox[origin=c]{90}{\textbf{Forced Registration}} & 
\parbox[c][3.1cm][t]{\linewidth}{%
\includegraphics[width=\linewidth]{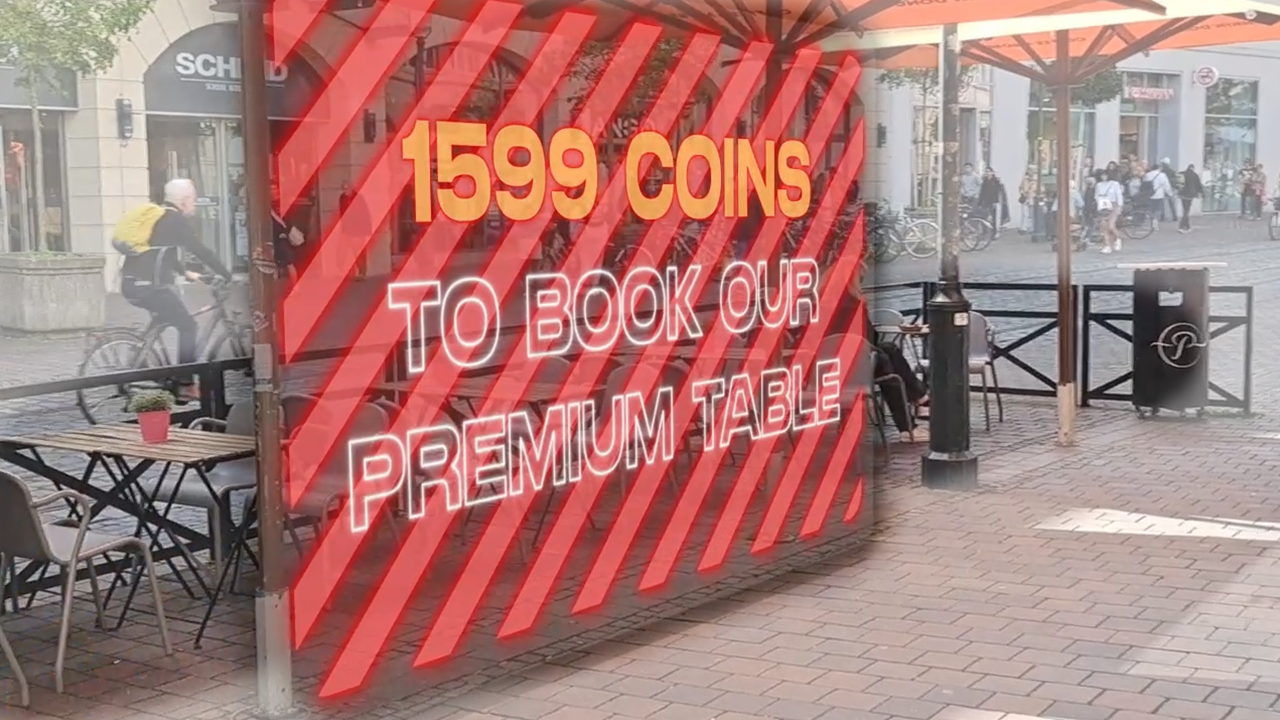} \newline 
A virtual wall was placed in front of the restaurant to prevent direct booking and see-through, displaying the message "1599 coins to book our premium table".} & 
\parbox[c][3.1cm][t]{\linewidth}{%
\includegraphics[width=\linewidth]{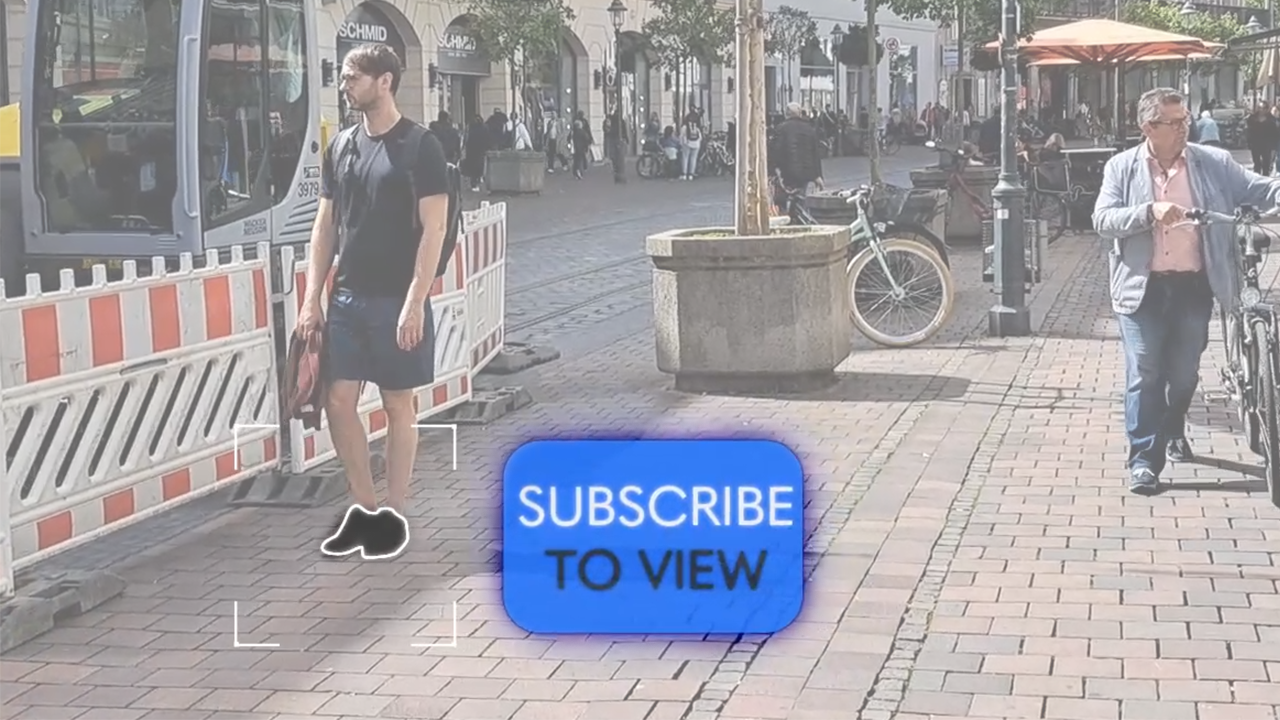} \newline 
The shoes were obscured with the message "Subscribe to View," forcing users to register to see them.} & 
\parbox[c][3.1cm][t]{\linewidth}{%
\includegraphics[width=\linewidth]{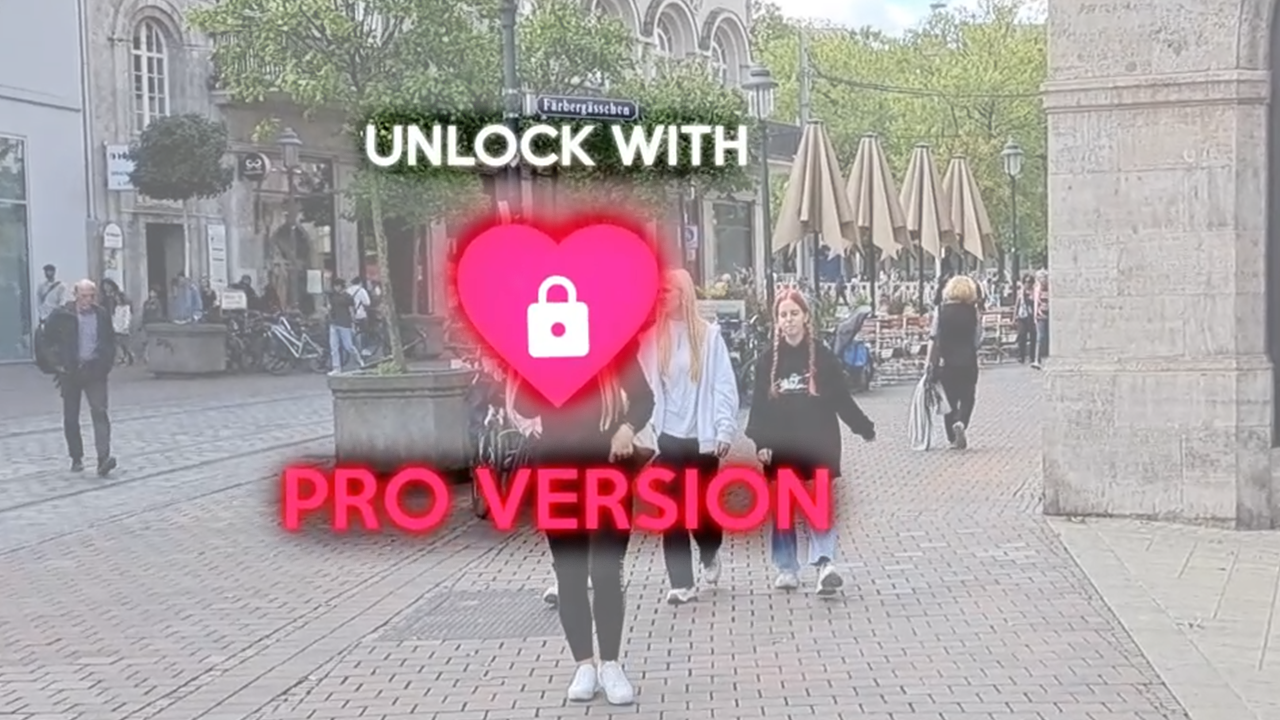} \newline 
The person's face was obscured with a heart and a lock, along with the text "Unlock with pro version," forcing users to upgrade to a paid subscription to view the face.} \\[15pt]

\rotatebox[origin=c]{90}{\textbf{Hiding Information}} & 
\parbox[c][3.1cm][t]{\linewidth}{%
\includegraphics[width=\linewidth]{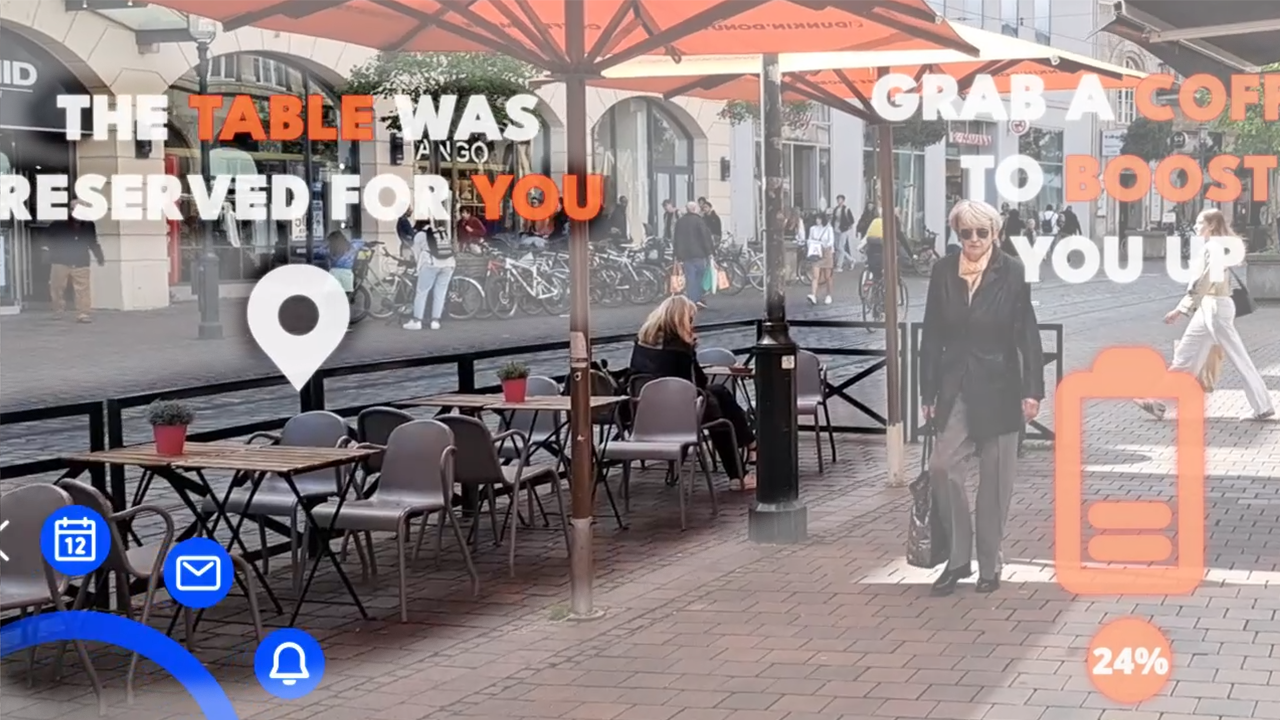} \newline 
Users were informed that a table at the restaurant had been booked for them without their consent.} & 
\parbox[c][3.1cm][t]{\linewidth}{%
\includegraphics[width=\linewidth]{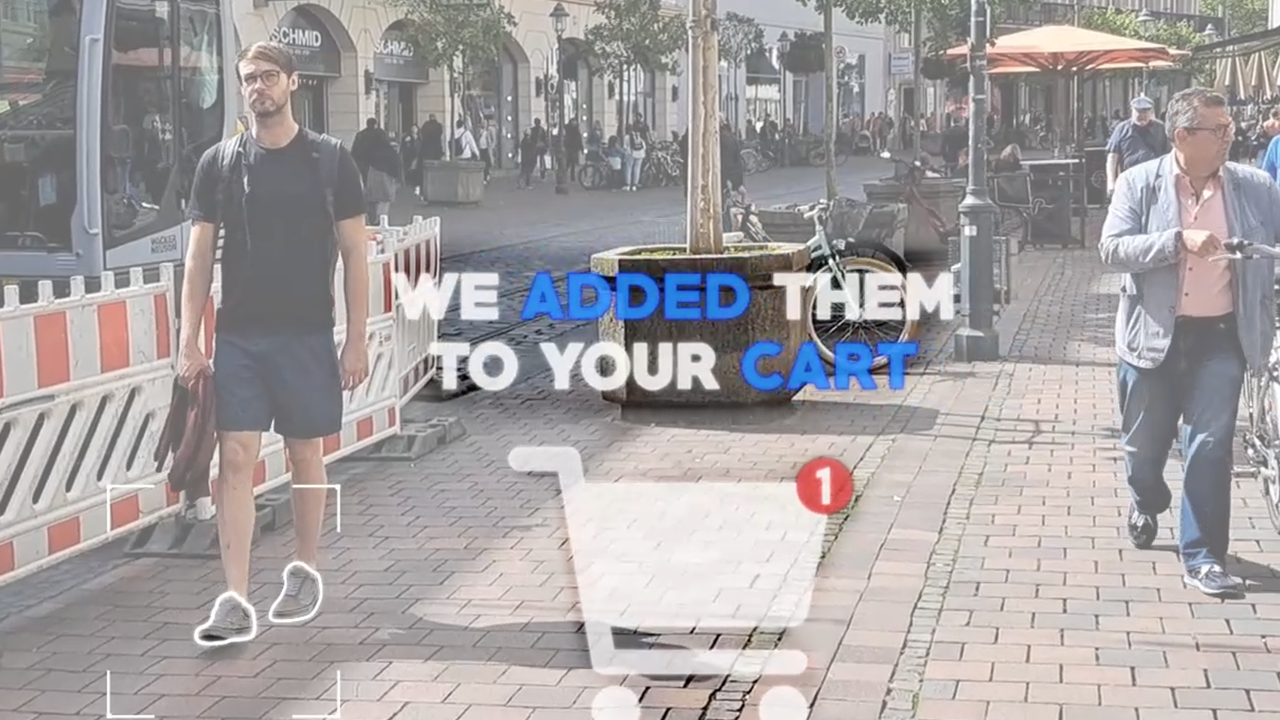} \newline 
Users were informed that the product had been added to the shopping cart without prior approval.} & 
\parbox[c][3.1cm][t]{\linewidth}{%
\includegraphics[width=\linewidth]{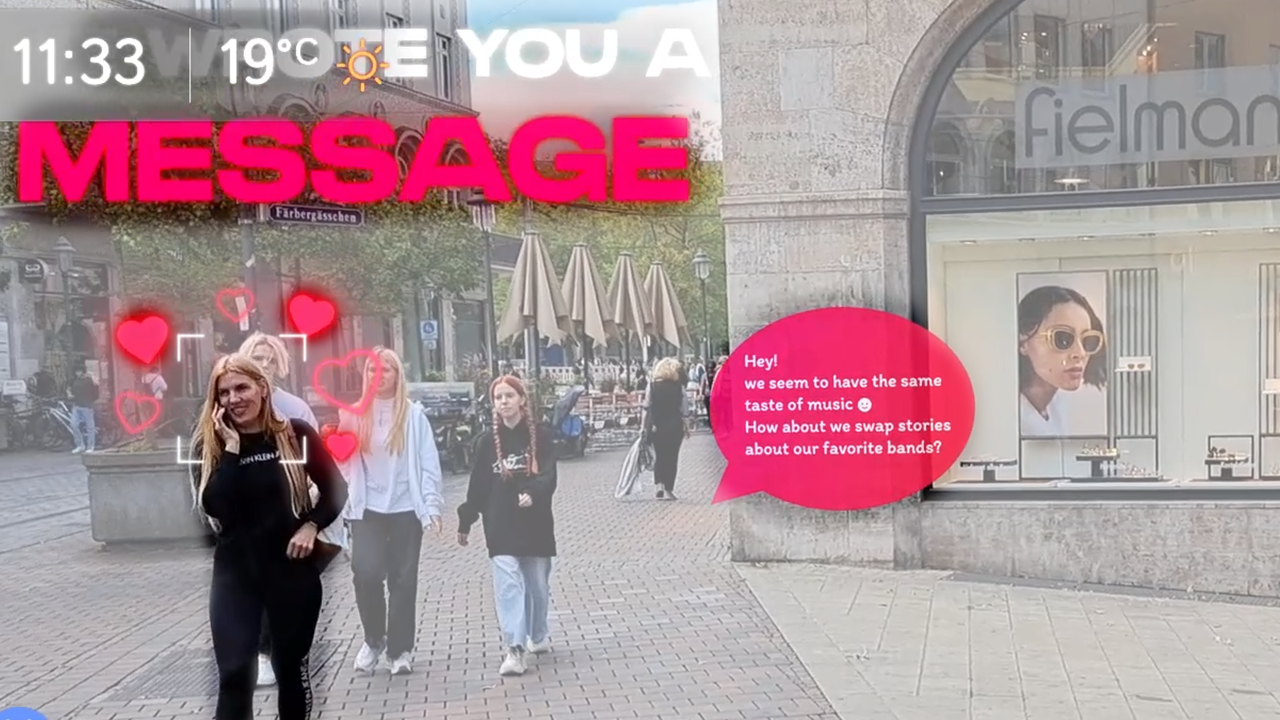} \newline 
Users were informed that a text message was pre-written to the person, which could be sent automatically by stepping into a predefined area.} \\[15pt]

\rotatebox[origin=c]{90}{\textbf{Urgency}} & 
\parbox[c][3.1cm][t]{\linewidth}{%
\includegraphics[width=\linewidth]{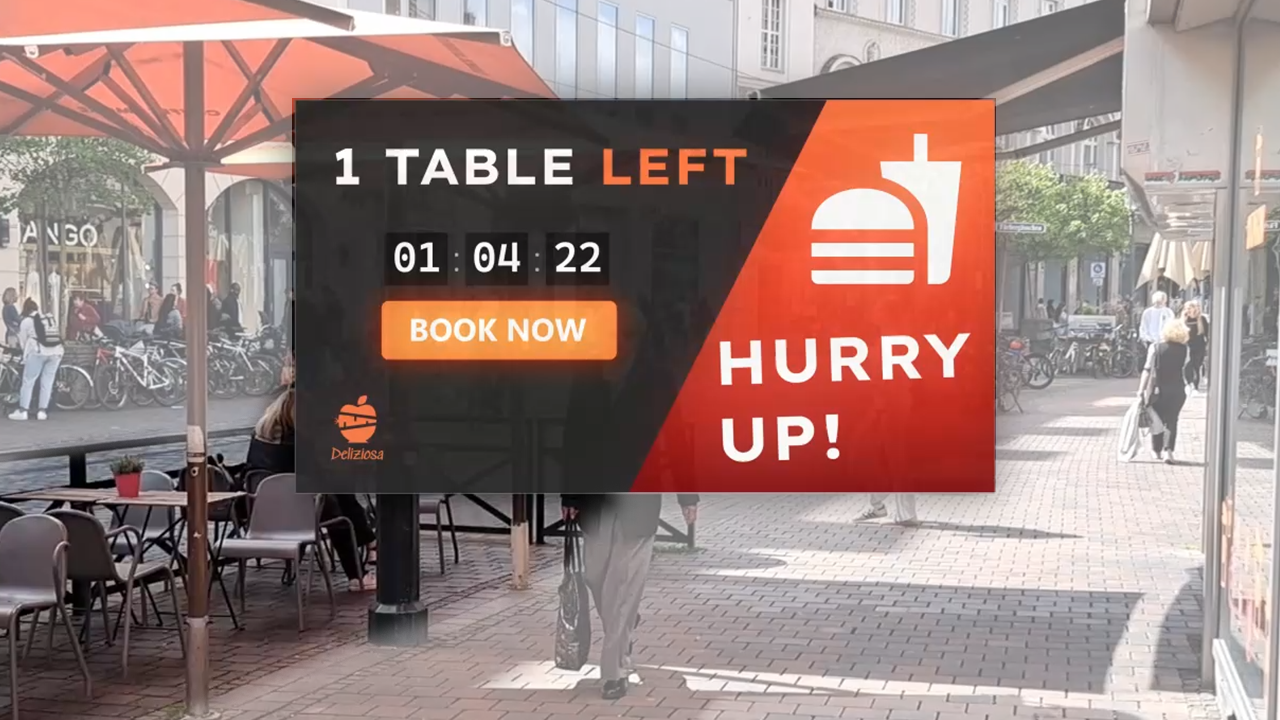} \newline 
To induce time pressure, we augmented a countdown timer and notified viewers that only one table was left to book, urging them to act quickly.} & 
\parbox[c][3.1cm][t]{\linewidth}{%
\includegraphics[width=\linewidth]{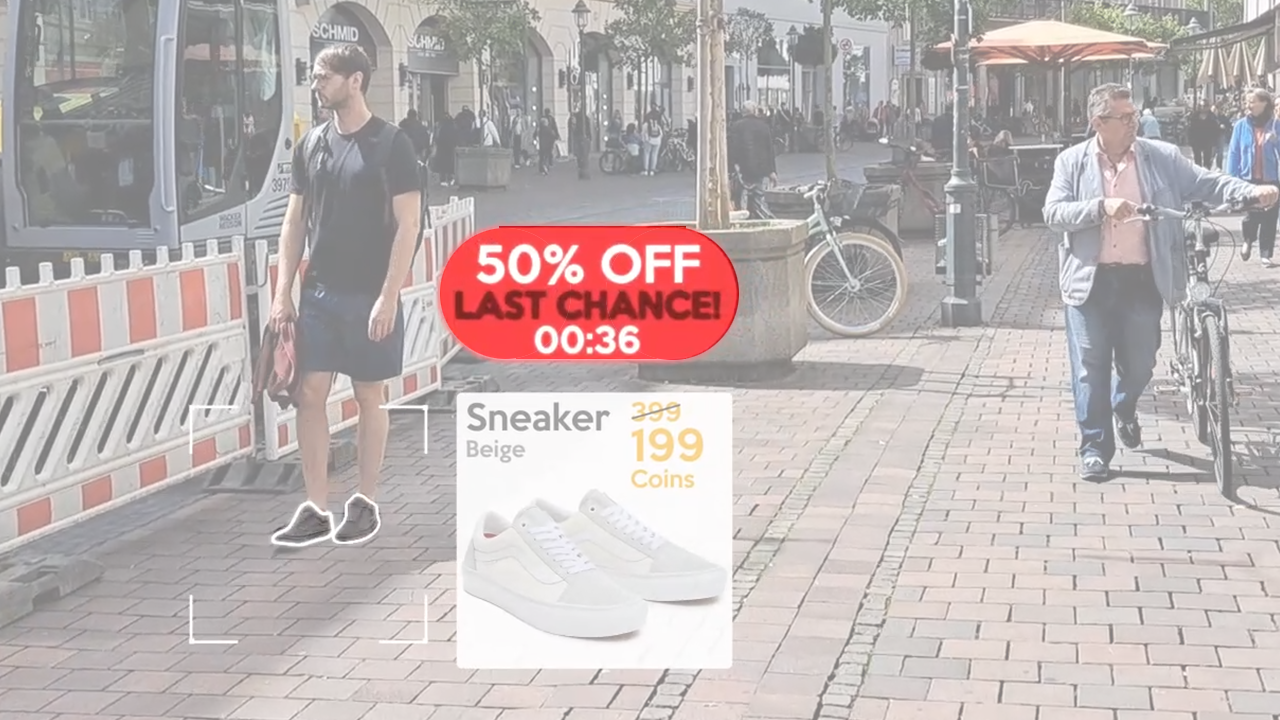} \newline 
We showcased the shoes with the augmented text "50\% off, last chance" and a countdown timer to prompt immediate purchasing.} & 
\parbox[c][3.1cm][t]{\linewidth}{%
\includegraphics[width=\linewidth]{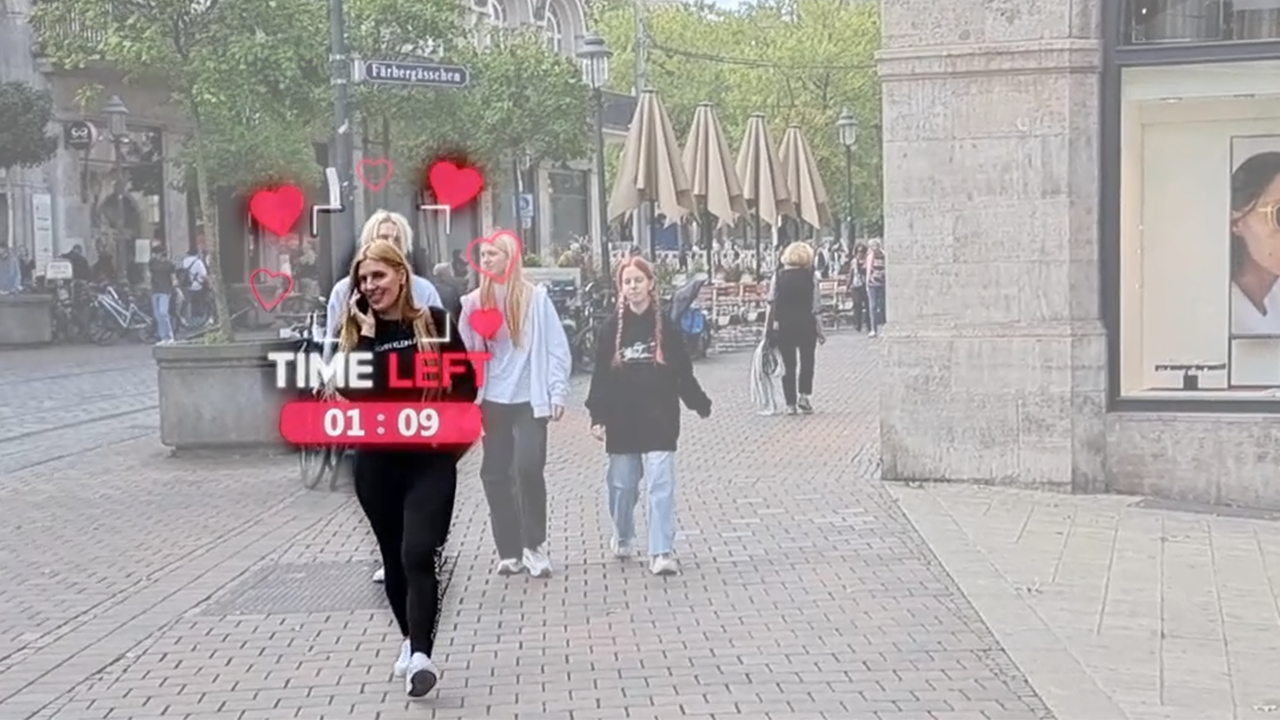} \newline 
The person’s face was augmented with animated hearts and a countdown timer, along with the text "time left," to create a sense of urgency.} \\[10pt]

\end{tabular}
\end{table*}

\subsubsection{Augmentation Targets}
We selected three specific targets: (1) a place (a restaurant in the pedestrian zone), (2) a product (shoes worn by a pedestrian), and (3) a person—to apply the four dark patterns, interpreting them within the context of their specific use or application type~\cite{Gray.2024}. These targets were chosen to represent clearly different categories, as originally described by \citet{greenberg_dark_2014}, and to explore how dark patterns from traditional GUI contexts can be adapted to MR. Each target captures a unique level of harm: augmentations applied to a person are typically more personal (see \citet{rixen_exploring_2021, bonner_when_2023}), whereas those targeting products or general environments may affect users differently. This approach enabled us to investigate whether the same dark pattern mechanism produces similar effects on users regardless of context, or if outcomes vary significantly across different targets of the physical world.
Each visual design element was rigorously discussed among the authors and refined before integration into the study, ensuring that the scenarios accurately reflect the implications of GUI dark patterns as defined by the established ontology in \citet{Gray.2024}.


Drawing inspiration from \citet{hyperreality}, we tailored each dark pattern to these specific targets in the MR environment with distinct fictional intentions. Thus, for the place (restaurant), we assumed that the restaurant would use MR to encourage viewers to book a table, reflecting real-world marketing strategies that impact the competitiveness of the restaurant industry~\cite{singh_impact_2024}. For the product (shoes), the goal was to promote the sale of the featured shoes, mirroring traditional advertising approaches adapted to MR and inspired by \citet{bonner_when_2023} who found negative effects when sponsored items (e.g., a logo of a company was augmented).
Lastly, for the person target, we envisioned a scenario inspired by dating apps expanding into MR~\cite{tinderverse}, prompting viewers to initiate interaction with the individual. This concept aligns with the potential for MR to display personal information, such as relationship status~\cite{kotyuk_machine_2012} or sexual orientation~\cite{jernigan_gaydar_2009}, as also highlighted by \citet{Rixen2}.



\subsection{Procedure}
Participants were recruited via \url{prolific.co} and pre-screened to live in the US to avoid cultural influences and because the US represents a major market for MR technology~\cite{MRTech-report}.
Each session began with a brief introduction, followed by signing the informed consent. We then introduced the participants to the scenarios through the following introductory text:

\begin{quote}
\noindent\textit{Imagine yourself walking through the pedestrian zone of a city center, equipped with a pair of Augmented Reality glasses. These glasses have the ability to display various visualizations and features from different apps in your surroundings. In the upcoming series of videos, each lasting 2 minutes, you will find yourself in this described situation. Your task is to fully immerse yourself in this scenario as if you were the protagonist and attentively observe these app visualizations.}
\end{quote}

\noindent Participants then watched the videos in randomized order. Subsequent to each video, participants completed the questionnaires outlined in \autoref{sec:measures}. On average, each session lasted $\approx$1~h. For their time and effort, the participants received a compensation of £6. This amount aligns with Prolific’s minimum recommended rate\footnote{\url{https://researcher-help.prolific.com/en/article/2273bd}, accessed Feb 2025} for a study of this duration and exceeds the minimum wage threshold for US participants, ensuring fair payment. Further, a background script ensured that videos played in full screen without the option to skip or replay, mirroring the real-time nature of MR experiences where content typically cannot be revisited. This design choice also prevented repeated exposure to the dark patterns that could have influenced the results. Because each video lasted only two minutes, the brief duration was expected to help maintain participants’ focus and reduce the risk of distractions causing them to miss important details. Additionally, the script forced the use of a FullHD (1080p) monitor or better to participate in the user study.

\subsection{Measurements}\label{sec:measures}
\citet{mildner_defending_2023} investigated the recognition of dark patterns in social media using five characteristics (asymmetry, covert tactics, deception, hiding information, and restriction) rated on a 5-point Likert scale based on \citet{mathur_dark_2019}. However, since this metric is not a validated scale, we refrained from using it. Instead, we employed the System Darkness Scale (SDS)~\cite{van_nimwegen_shedding_2022} to measure the impact of dark patterns on users. This scale is designed "\textit{as a validated tool to identify in how far a system or service has incorporated 'dark mechanisms'\,}"\cite[p. 1]{van_nimwegen_shedding_2022} and provides an impression of user perceptions of dark patterns in a system. While initially applied to web shops, we consider SDS equally relevant for evaluating user perceptions of dark patterns in MR. Its emphasis on capturing how users experience and recognize manipulative design makes it a promising tool for our context.

Additionally, we used reactance to determine whether participants perceived their freedom of choice to be threatened by the dark patterns. In HCI, reactance is defined as the resistance that individuals feel when their freedom is perceived to be threatened~\cite{Ehrenbrink.2020}, which is consistent with psychological models that behavior change interventions limit an audience's freedom~\cite{Dillard.2005, Rains.2013}. Since dark patterns, by definition, "\textit{[...] make you do things you did not want to do [...]}"~\cite{brignull-2023}, we argue that they inherently reduce the user's freedom of choice. Therefore, reactance is a well-suited measure for assessing dark patterns' impact on users.

Further, we evaluated the participants' comfort level using a single-item question. Inspired by \citet{rixen_exploring_2021}, we asked participants to rate their comfort with the augmentations using the question, "Please rate how comfortable you felt with the augmentations," on a 5-point Likert scale (from 1 - very uncomfortable to 5 - very comfortable). Their research showed that user comfort can vary significantly depending on which body parts are augmented. Hence, we assume that user comfort may also vary based on the augmentation targets of different dark patterns in MR.

Lastly, we measured the intention to use a system incorporating the dark patterns shown. This measurement is based on the Technology Acceptance Model (TAM) developed by \citet{venkatesh_theoretical_2000}. The TAM framework helps to understand user acceptance of technology by evaluating perceived usefulness and ease of use. For our study, we included the items from the intention-to-use-subscale, rated on a 7-point Likert scale: "Assuming I have access to the system, I intend to use it" and "Given that I have access to the system, I predict that I would use it." This metric allowed us to measure participants' intentions toward using an MR system despite the presence of dark patterns, providing insight into how they might influence user adoption and acceptance.

After all conditions, we used the Immersion subscale of the Technology Usage Inventory (TUI)~\cite{kothgassner2013technology} to confirm that adequate immersion was achieved. Further, participants provided open feedback via a text field asking, "Do you have any comments regarding the different augmentations you saw?".
All study question items are detailed in \autoref{app:usedquestionaires}.

\subsection{Results}\label{sec:Results}
The user study was conducted as a 4 $\times$ 3 design + 1 baseline condition. Hence, before the two-factorial analyses of \textit{Dark Pattern} $\times$ \textit{Augmentation}, we conducted a baseline comparison by comparing all conditions with the baseline. Before performing any statistical tests, we used the Shapiro-Wilk test~\cite{Shapiro-Wilk} to check the normal distribution of the data. The test found that all dependent variables did not follow a normal distribution. Hence, for the baseline comparison, we used Friedmann's ANOVA with the Durbin-Conover post-hoc test using the Holm correction. For the subsequent two-factorial analysis, we used the aligned rank transformation (ART) for non-parametric factorial analyses for repeated measures~\cite{wobbrock_aligned_2011}.
We conducted our analyses in R version 4.3.1. \autoref{app:descriptive} shows detailed descriptive statistics. 

\subsubsection{Participants}\label{sec:participants}
For the online study, we recruited 79 participants but excluded five due to incorrect attention checks, resulting in a final sample of 74. These seven nonsensical attention checks occurred randomly in the survey, including questions such as ``\textit{I swim across the Atlantic Ocean every day to get to work}'' or ``\textit{I was born in the 13th century.}'' We aligned these attention checks with Prolific's policy\footnote{\url{https://researcher-help.prolific.com/en/article/fb63bb}, accessed Apr 2025} and rejected participants who failed at least two of them. The ages of participants ranged from 18 to 74 years (\m{35.27}, \sd{11.31}). Of these, 34 identified as male, 37 as female, and three as non-binary. Regarding education, 30 participants held high school degrees, 22 had bachelor's degrees, 13 had master's degrees, 6 had doctoral degrees, and 3 had completed apprenticeships. Further, current occupations were distributed as follows: 35 employed, 11 self-employed, 12 full-time students, 9 unemployed/job-seeking, 2 retired, and 5 selecting “other” (e.g., homemaker). After all conditions, we asked about the participants' immersion~\cite{kothgassner2013technology} during the video study, which was rated medium to high (\m{16.28}, \sd{6.58}; minimum possible: 4, maximum possible: 28). 

\subsubsection{Reactance}

\begin{figure*}[ht!]
\centering
\small
    \begin{subfigure}[c]{0.47\linewidth}
        \includegraphics[width=\linewidth]{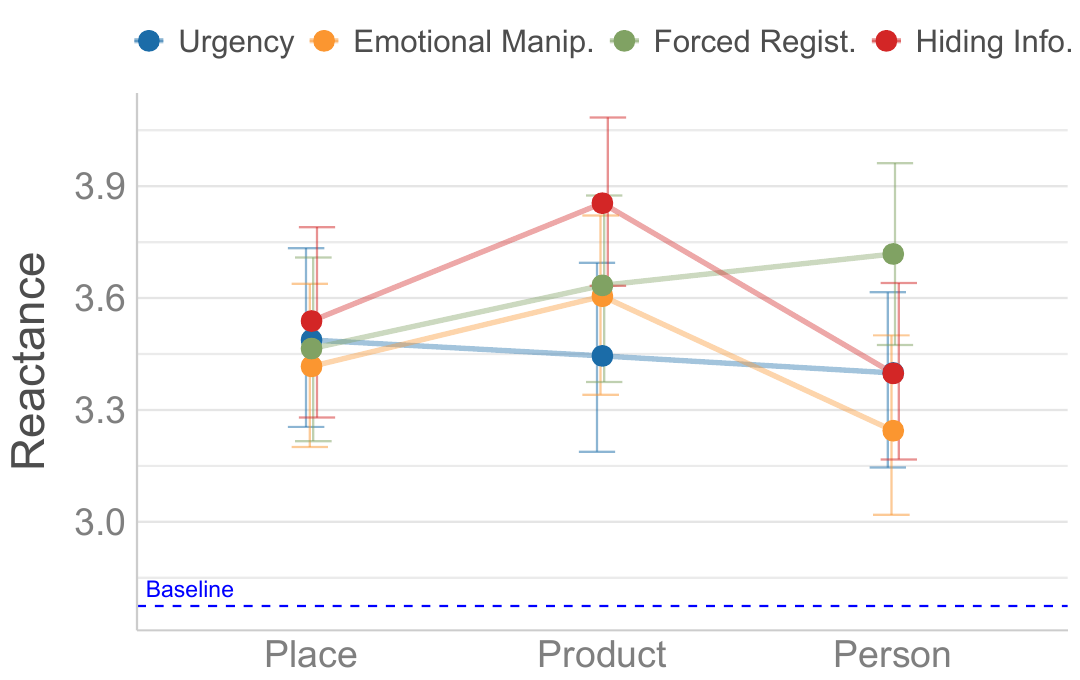}      \caption{Reactance~\cite{Ehrenbrink.2020}}\label{fig:reactance_interaction}
        \Description{Interaction effect of reactance including the rating of the baseline, as a dotted line below the ratings for the dark pattern scenarios}
    \end{subfigure}
    \hspace{0.5cm}
    \begin{subfigure}[c]{0.47\linewidth}
        \includegraphics[width=\linewidth]{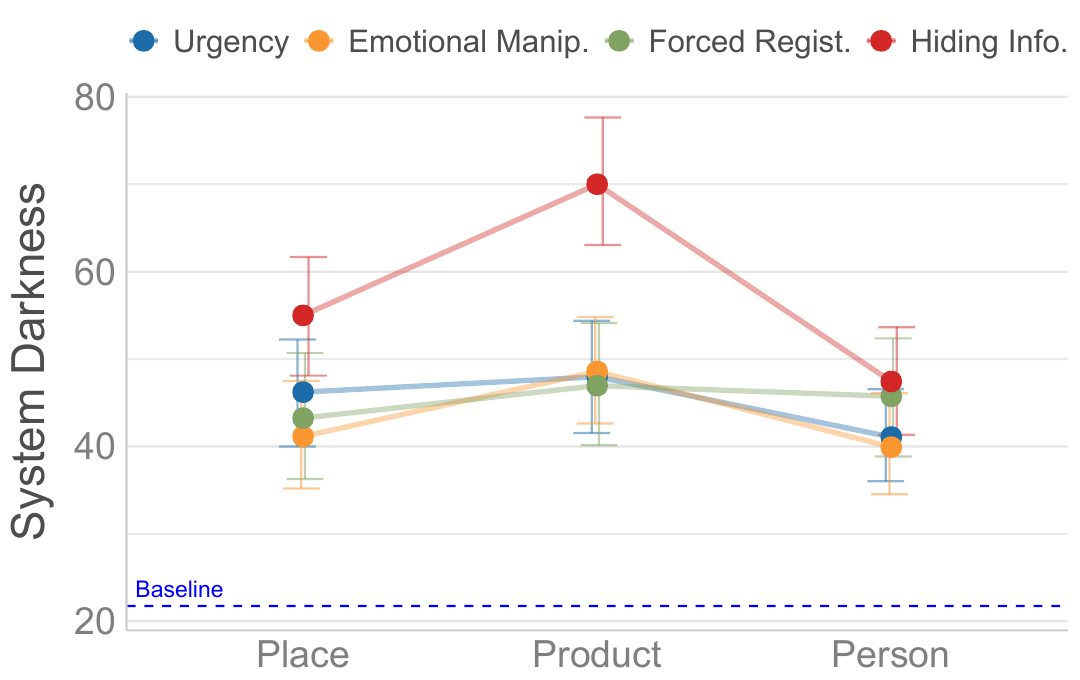}
        \caption{System darkness scale~\cite{van_nimwegen_shedding_2022}}\label{fig:systemDarkness_interaction}
        \Description{Interaction effect of the system darkness scale including the rating of the baseline, as a dotted line below the ratings for the dark pattern scenarios}
    \end{subfigure} 
   \caption{Interaction effects on reactance and System darkness scale}
   \Description{This plot shows the interaction effects on reactance and the System Darkness Scale}
\end{figure*}

The baseline comparison via Friedmann test (\chisq(12) = 97.67, \pminor{0.001}, $\hat{W}=0.11$) showed that after the post-hoc tests, the reactance of all conditions was significantly higher than the baseline condition (\m{2.77}, \sd{0.85}).
For the two factorial ART, we found a significant main effect for both \textit{Dark Pattern} (\F{3}{219}{8.11}, \pminor{0.001}, $\eta^2$ = 0.10) and \textit{Augmentation Target} (\F{2}{146}{6.71}, \p{0.002}, $\eta^2$ = 0.08). A significant interaction (\F{6}{438}{4.43}, \pminor{0.001}, $\eta^2$ = 0.06) shows the interplay between \textit{Dark Pattern} $\times$ \textit{Augmentation Target} (see \autoref{fig:reactance_interaction}). For \textit{Emotional or Sensory Manipulation} and \textit{Hiding Information}, reactance followed a similar pattern across the three augmentation targets, with \textit{Product} showing the highest reactance, while \textit{Place} and \textit{Person} showed similar, lower scores. In contrast, for \textit{Forced Registration}, reactance increased from \textit{Place} (lowest) to \textit{Person} (highest).

\subsubsection{System Darkness Scale}
For the SDS, the post-hoc tests of the baseline comparison, after Friedmann's test (\chisq(12)=174.85, \pminor{0.001}, $\hat{W}=0.20$), showed that all conditions were significantly higher than the baseline condition (\m{21.76}, \sd{21.64}).
For the two factorial ART, we further found a significant main effect of \textit{Dark Pattern} on the System Darkness Scale (\F{3}{219}{29.82}, \pminor{0.001}, $\eta^2$ = 0.29) and of \textit{Augmentation Target} (\F{2}{146}{21.27}, \pminor{0.001}, $\eta^2$ = 0.23). Additionally, a significant interaction effect of \textit{Dark Pattern} $\times$ \textit{Augmentation Target} was significant (\F{6}{438}{6.50}, \pminor{0.001}, $\eta^2$ = 0.08) as shown in \autoref{fig:systemDarkness_interaction}. All ratings for the dark patterns, except for \textit{Hiding Information}, are similar, with little variation among the three augmentation targets. However, \textit{Hiding Information} stands out, showing significantly higher SDS ratings for \textit{Product} compared to the other dark patterns and the other augmentation targets.

\subsubsection{Comfort}\label{sec:results_comfort}
\begin{figure*}[ht!]
\centering
\small
    \begin{subfigure}[c]{0.47\linewidth}
        \includegraphics[width=\linewidth]{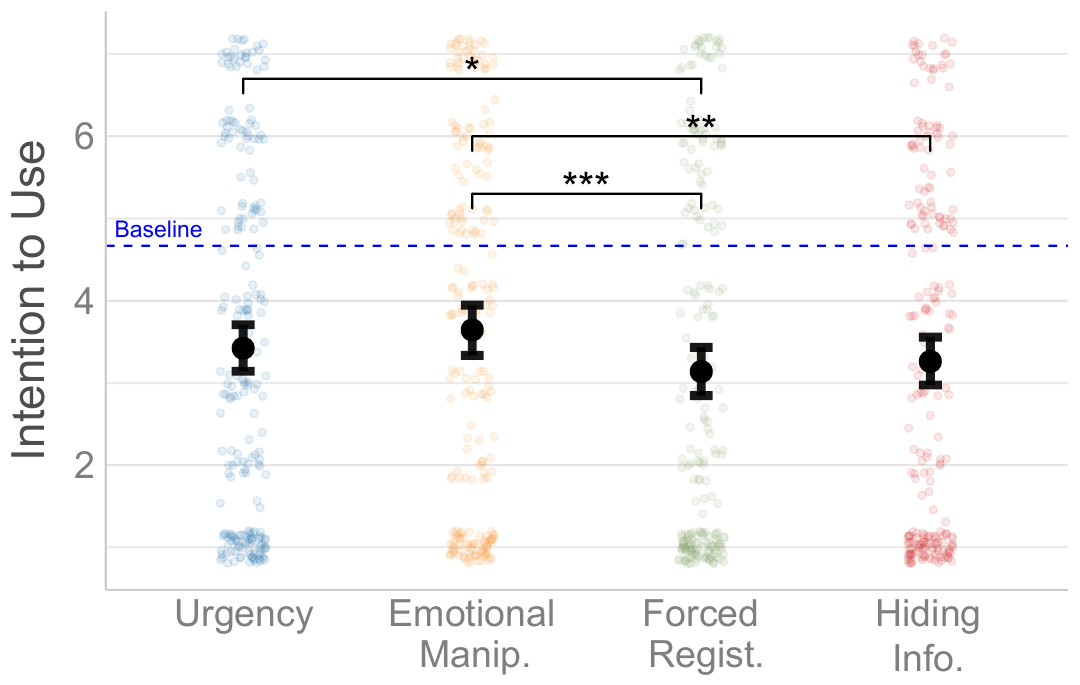}
        \caption{Intention to use~\cite{venkatesh_theoretical_2000}}\label{fig:intentiontouse_main}
        \Description{This plot presents the main effects of different dark patterns on the intention to use an MR system. Emotional or Sensory Manipulation received the highest ratings, while Forced Registration resulted in the lowest. The baseline rating, shown as a dotted blue line, is positioned above all the dark pattern ratings}
    \end{subfigure}
    \hspace{0.5cm}
    \begin{subfigure}[c]{0.47\linewidth}
        \includegraphics[width=\linewidth]{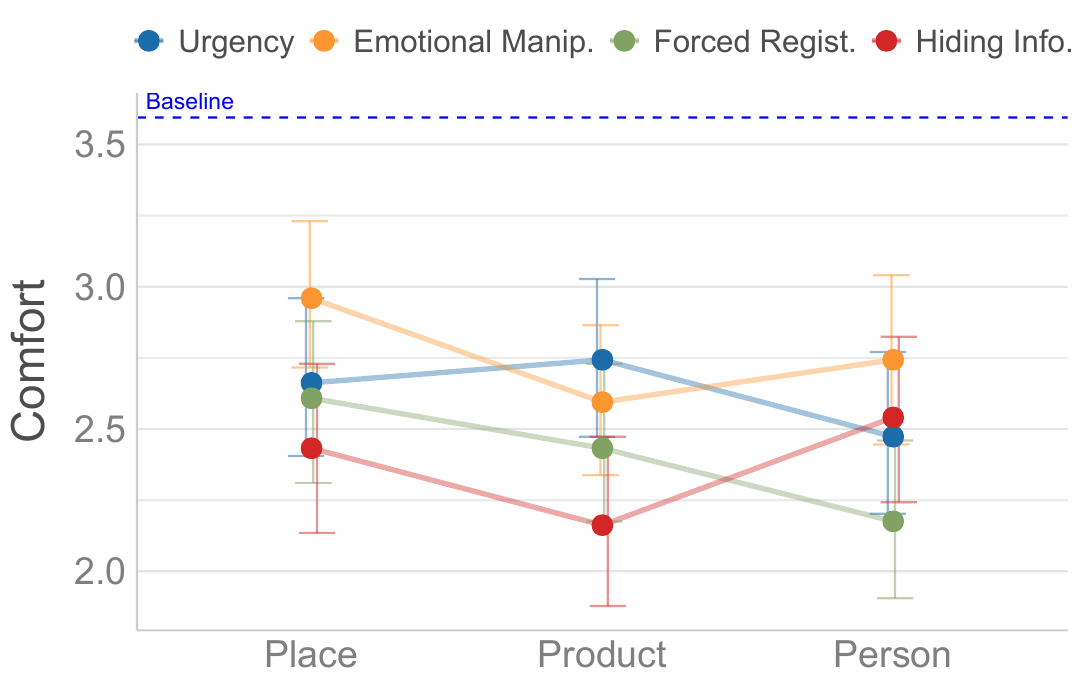}  \caption{Comfort~\cite{rixen_exploring_2021}}\label{fig:comfort_interaction}
        \Description{Interaction effect of perceived level of comfort including the rating of the baseline, as a dotted line above the ratings for the dark pattern scenarios}
    \end{subfigure} 
   \caption{Interaction effects on Comfort and main effect on intention to use}\label{fig:comfortandintentiontouse}
   \Description{These plots show the main effect on the intention to use and the interaction effect on the rated comfort level}
\end{figure*}

The baseline comparison for the rated comfort was significant (\chisq(12)=145.89, \pminor{0.001}, $\hat{W}=0.16$). Hence, the post-hoc test revealed that the baseline (\m{3.59}, \sd{1.01}) condition was significantly higher rated compared to all other conditions. The subsequent two factorial ART found significant main effects for \textit{Dark Pattern} (\F{3}{219}{14.93}, \pminor{0.001}, $\eta^2$ = 0.17) and \textit{Augmentation Target} (\F{2}{146}{3.56}, \p{0.031}, $\eta^2$ = 0.05). Further, there was also a significant interaction effect of \textit{Dark Pattern} $\times$ \textit{Augmentation Target} (\F{6}{438}{4.76}, \pminor{0.001}, $\eta^2$ = 0.06) depicted in \autoref{fig:comfort_interaction}. Similar to the reactance, the rating for \textit{Emotional or Sensory Manipulation} and \textit{Hiding Information} follow a similar pattern across the three augmentation targets, with \textit{Product} yielding the lowest comfort level. The pattern is inverted for the \textit{Urgency} dark pattern as here the \textit{Product} yielded the highest score among the three augmentation targets. For \textit{Forced Registration}, there is a decrease of comfort from \textit{Place} (highest) to \textit{Person} (lowest).

\subsubsection{Intention to Use}
The baseline comparison via post-hoc tests after the Friedmann test ((\chisq(12)=82.07, \pminor{0.001}, $\hat{W}=0.09$)) showed that the baseline condition (\m{4.67}, \sd{2.01}) was significantly higher in terms of the intention to use compared to all other conditions. The subsequent two factorial analyses via ART revealed a significant main effect of \textit{Dark Pattern} on intention to use (\F{3}{219}{7.92}, \pminor{0.001}, $\eta^2$ = 0.10). A post-hoc test showed significant differences between \textit{Emotional or Sensory Manipulation} and \textit{Forced Registration} (\pminor{0.001}), \textit{Emotional} and \textit{Hiding Information} (p\textsubscript{adj}=0.01) as well as between \textit{Urgency} and \textit{Forced Registration} (p\textsubscript{adj}=0.04). More detailed, the \textit{Emotional or Sensory Manipulation} dark pattern yielded the highest intention to use (\m{3.64}, \sd{2.24}) while the lowest intention to use was rated for the \textit{Forced Registration} dark pattern (\m{3.14}, \sd{2.21}). The main effect is depicted in \autoref{fig:intentiontouse_main}.

\subsubsection{Open Feedback}
For the open feedback, we did not conduct a thematic analysis. Instead, we present anecdotal feedback and participants' opinions. 44 participants provided relevant (i.e., not "thanks") open feedback. 
The qualitative feedback from participants in our study revealed a range of negative impacts primarily affecting user experience and well-being.
Participants frequently reported frustration and annoyance due to certain dark patterns. For instance, P2 highlighted patterns that "\textit{blocked out products or faces actively disgusted me.}" A similar sentiment regarding the Forced Action dark pattern was expressed by P7 and P33. Especially, P68 felt that the augmentations blocking his view "\textit{felt particularly dystopian to me}". Additionally, sneaking a product into the basket was perceived as disruptive, as it happened without the consent of the viewer (P56).
Several participants found the \textit{Forced Registration} dark pattern particularly bothersome. For instance, four participants (P71, P31, P14, P10) said they did not like being forced to do something during daily life. Further aversion was especially noted in scenarios involving augmentations that targeted the person. For example, P16 commented that such augmentations made him "\textit{[..] think on how dangerous it can be to have that information right there [at the MR glasses] and then in terms of harassment [..]}". This concern has been previously discussed by \citet{freeman_disturbing_2022}, whose research on VR chat concluded that immersive technologies like VR and MR may give rise to new forms of harassment. On the other hand, P36 and P70 mentioned that these augmentations could potentially improve social interactions, provided the augmented person gave consent. In general, five participants (P66, P2, P37, P38, P33) noted that the only useful augmentations were those used in the baseline condition, indicating a preference for simpler, less intrusive augmentations. However, concerns were raised about the physical impact of large augmentations. P44, P62, and P39 observed that some augmentations were so large that they blocked out the real world, creating potential hazards. P22 specifically mentioned the dangers posed by such blocking, highlighting the safety risks associated with overly intrusive augmentations. \\
In conclusion, the feedback underscores the predominantly negative effects of dark patterns in MR environments, including frustration, annoyance, and potential privacy hazards, which aligns with findings of prior studies outside MR contexts~\cite{luguri_shining_2021,Gray.2018}. However, there were also positive comments. For instance, P69 mentioned that "\textit{[..] it was a positive experience, and I would definitely purchase an MR headset.}"

\section{Discussion}




This work explored the impact of dark patterns in MR when applied to different real-world targets. Hence, we conducted a video-based online study with N=74 participants. Our 13 videos depicting a city walkthrough from the perspective of an MR glasses user constituted a 4 $\times$ 3 + 1 baseline factorial design. The independent variables were four dark patterns (\textit{Emotional or Sensory Manipulation}, \textit{Forced Registration}, \textit{Hiding Information}, and \textit{Urgency}) aligned with the meso-level patterns identified by \citet{Gray.2024}. These dark patterns were augmented to three different targets: a place (a restaurant in the pedestrian zone), a product (shoes of a pedestrian), and a person. This section explores the implications of our findings and discusses the potential impacts and risks that may arise. 

\subsection{Impacts of Dark Patterns on Targets in Mixed Reality}
All investigated dark patterns in the MR environment were clearly recognized by users, as evidenced by significantly higher SDS ratings compared to the baseline scenario without dark patterns. These results align with the 87.5\% correct identification rate of dark patterns in social media applications~\cite{mildner_defending_2023}, suggesting that, regardless of the context, users are aware of the disruptive appearance of dark patterns. However, the immersive nature of MR could make falling for these even more pronounced in the future, in the same way that dark patterns have gained traction on social media~\cite{mildner_defending_2023}.

To shift the research focus from qualitative (e.g., \cite{hadan_deceived_2024, tseng_dark_2022, krauss_what_2024}) towards quantitative methods, we employed concrete visualizations of dark patterns in MR and utilized quantitative metrics (reactance~\cite{Ehrenbrink.2020}, SDS~\cite{van_nimwegen_shedding_2022}, intention to use~\cite{venkatesh_theoretical_2000} and comfort~\cite{rixen_exploring_2021}) to answer our initial RQ (see \autoref{sec:intro}). We found that all dark patterns significantly impact our quantitative measures negatively, regardless of whether they target a place, product, or person.
While every dark pattern caused heightened reactance and reduced comfort levels compared to the baseline, the severity of the negative impact intensified when the dark pattern involved personal interference or financial manipulation. Here, two combinations of target and dark pattern were identified as particularly negative, eliciting increased reactance and a low comfort level: \textit{Forced Registration}, by obscuring a \textit{person}'s face and \textit{Hiding Information}, by sneaking the \textit{product} into the user’s shopping basket without their consent. 

Participants reacted strongly to the augmentation of these dark patterns onto the target person and product in terms of heightened reactance and reduced comfort, implying their sensitivity involving monetary intentions (buying products) and personal interference (hiding other people's faces). This sensitivity to personal information in MR is supported by \citet{Rixen2}, who found that displaying confidential personal information such as relationship status or political views in MR decreases users' comfort. Our study extends these findings by implying that obscuring visual personal information, such as pedestrians' faces, also impacts comfort negatively. However, this reasoning does not fully explain why augmenting a person with \textit{Emotional or Sensory Manipulation} resulted in relatively low reactance and high comfort. In this dark pattern, the person was surrounded by animated hearts alongside the text "She likes you" to create emotional manipulation. Although this text might be seen as the person's privacy violation, participants might have perceived it as flattering, which leads to a more positive overall impression~\cite{lee_flattery_2008}. Thus, our results show that this flattering text yielded a lower reactance, and higher comfort than blocking someone’s face. Ironically, this makes \textit{Emotional or Sensory Manipulation} potentially more dangerous as users recognize it as manipulative, yet the positive framing may diminish their resistance towards the dark pattern. \\
Further, this study focused only on the perceiver’s perspective (the one wearing the MR glasses), not on how the target might feel. Prior work suggests that targets experience more discomfort when their personal information is revealed compared to the perceiver~\cite{Rixen2}. This difference likely contributed to the lower reactance observed for \textit{Emotional or Sensory Manipulation} augmented to human targets.
Likewise, the strong negative response in terms of comfort to \emph{Hiding Information} with products suggests that users quickly recognize the financial consequences, consistent with dark patterns found on shopping websites where unwanted items are added to the cart. This sensitivity was further confirmed by a high SDS rating for \emph{Hiding Information} with products. As the SDS scale is particularly designed "[...]\textit{to identify in how far a system or service has incorporated 'dark mechanisms'\,}"\cite[p. 1]{van_nimwegen_shedding_2022}, the high SDS rating imply a sensitive perception of deception when users sense financial consequences.

It is important to note that the prominence of these patterns may have made them more noticeable in this user study, mainly since participants were likely unfamiliar with similar manipulative tactics in MR settings. However, simply recognizing a dark pattern does not necessarily prevent it from being effective. For instance, infinite scrolling on smartphone applications is a well-known dark pattern~\cite{mongeroffarello2022towards}. Despite awareness~\cite{ko2015nugu}, people often continue to scroll endlessly through social media feeds, frequently experiencing regret afterwards~\cite{Mildner.2021, Cho.2017}.

\subsection{Rethinking Dark Pattern Categories in Mixed Reality}
The interaction effects between augmentation targets and dark patterns reveal that the impact of these targets on the users is similar for the dark patterns \textit{Emotional or Sensory Manipulation} and \textit{Hiding Information}, particularly in terms of rated reactance and comfort. This suggests that, regardless of whether the dark pattern is applied to a place, product, or person, these two patterns produce comparable user responses. However, earlier classifications (e.g., \cite{Gray.2018}) distinguished these dark patterns as separate categories based on their manipulation strategies. This distinction persisted until more recent classifications~\cite{Gray.2024}.  Nevertheless, our findings suggest that, at least in MR environments, the impact on users for the dark patterns \textit{Emotional or Sensory Manipulation} and \textit{Hiding Information} is comparable. While we cannot generalize this similarity to other media beyond MR, it raises an important consideration: current classifications of dark patterns are primarily based on deceptive design techniques rather than their actual impact on users. Therefore, we recommend that future classifications should also consider the quantitative impact of dark patterns on users. This approach could lead to new groupings and a more accurate classification of dark patterns based on their effects rather than just their manipulation techniques. 

This impact-based classification could be based on the metrics that we used, such as comfort~\cite{rixen_exploring_2021}, intention to use~\cite{venkatesh_theoretical_2000}, reactance~\cite{Ehrenbrink.2020}, and the SDS~\cite{van_nimwegen_shedding_2022}. Yet, the SDS was inherently designed to capture how users perceive dark patterns in webshops. Although its applicability to other media is not established, it is currently the only validated tool to assess dark patterns. Therefore, we argue that the SDS may also provide valuable insights into the impact of dark patterns in MR environments. 

Nevertheless, there is no validated scale for measuring the impact of dark patterns across emerging media, such as MR yet. We therefore recommend that future work should move beyond simply categorizing deceptive UI techniques and instead develop and validate metrics that capture how these manipulations actually affect users.
This might help categorize dark patterns based on \textit{how} they affect users, rather than \textit{what} designers do to deceive UIs.


\subsection{The Dystopian Potential of Dark Patterns in Mixed Reality and Mitigation Strategies}

Due to the immersive nature of MR, there is a substantial risk that users could become accustomed to manipulations~\cite{hadan_deceived_2024}. Unlike traditional interfaces (e.g., smartphones or computers), which have the display as a natural border between the virtual and physical world, MR directly integrates virtual elements into the user's physical environment, making it more difficult to distinguish between reality and dark patterns~\cite{krauss_what_2024}. 
For instance, our study highlighted participants' concerns about virtual objects obscuring real-world hazards. This concern was hypothesized by \citet{hadan_deceived_2024}, who argue that obscuring reality in MR can amplify adverse effects compared to traditional, website-based dark patterns. Removing objects in VR was already demonstrated by \citet{tseng_dark_2022}; however, MR now has advanced to the point where it can also remove objects from the real world, known as \emph{Diminished Reality}. This concept is defined as "\textit{a set of methodologies for concealing, eliminating, and seeing through objects in a perceived environment in real time to diminish reality}"~\cite[p. 1]{mori_survey_2017}.  
Some works applied diminished reality to remove entire objects considered irrelevant, thereby reducing visual distractions (e.g., removing a person~\cite{kimDonBotherMe2020}, vehicles~\cite{colley_feedback_2022}, or buildings~\cite{colley_feedback_2022}). While useful, this capability could be exploited in MR to create dark patterns, such as hiding competing brands in stores to limit consumer choice. Considering the potential seamless integration of dark patterns into MR environments and the growing ability to add, alter, or hide elements of reality, it is not difficult to envision a dystopian future where users are constantly subjected to subtle, immersive manipulations. 

\subsubsection{Educational and Policy Implications}
As highlighted by the Collingridge-Dilemma~\cite{collingridge_social_1982}, we cannot fully understand how MR will evolve without allowing it to evolve naturally. If we intervene too early, we may miss its potential development, yet once mature, interventions become challenging. This research provides only an early glimpse into how dark patterns might manifest in MR, leaving open the question of how these manipulative design techniques will develop over time. However, the rapid emergence of immersive technologies, illustrated by new devices such as Meta Orion~\cite{metaOrion}, indicates that MR is already becoming a reality. In light of these findings, we argue that regulations have to be put into place before the technology can be exploited, protecting users in similar ways as current regulations, addressing dark patterns in traditional interfaces~\cite{DSA, FTC}.

We also argue that designers and practitioners should follow a strand of good practice and value-sensitive design guidelines~\cite{doorn_value_2013} to keep users' agency high and avoid dark patterns, for instance, by making any manipulation on perceived content transparent beforehand. In this vein, \citet{gray_scaffolding_2023} worked together with practitioners to develop ethics-focused design considerations that can be used in design stages to mitigate implementations of deceptive designs. Summarized on a supplementary website~\footnote{\href{https://everydayethics.uxp2.com/methods/}{https://everydayethics.uxp2.com/methods/}, accessed Apr 2025.}, these methods integrate ethical values, including empathy, inclusiveness, and accessibility.  
Importantly, \citet{chivukula_wrangling_2023} point to systemic issues between a practitioner's values and their organization's goals that might overrule a practitioner's concerns, as \citet{gray_ethical_2019} describe ethical mediators to limit these problems and foster ethical practice. 
These findings are also relevant in MR context, where designers can adopt ethical design approaches, embedding transparency and user agency for MR interfaces. For instance, any feature that alters or obscures reality should come with clear notifications and easy-to-use control options. Similar to accessibility checkers that ensure designs meet inclusion standards, a “dark pattern checker” could be developed to alert designers to potentially deceptive designs in MR. 
Moreover, conducting workshops among researchers and designers (e.g., by \citet{lukoff_what_2021}) is an essential first step to enhancing awareness of dark patterns in the community. Although such workshops may initially reach only a limited audience, increasing awareness will empower designers to integrate ethical guidelines into MR systems before the technology becomes ubiquitous. \citet{eghtebas_co-speculating_2023} highlighted the threat posed by dark patterns in AR and emphasized the need for proactive countermeasures. 
Building on their suggestions and our findings on the negative effects of dark patterns in MR, we urge stakeholders to adopt countermeasures like automated detection tools or an “ethical mode” (e.g., a user-selectable option that neutralizes or flags suspicious MR behavior). 











\subsection{Limitations and Future Work}
This video-based online study has limitations in external validity and transferability due to using online videos instead of an actual MR application. However, our approach avoided hardware-centric barriers and allowed participants without MR or VR headsets to participate in this study, ensuring a higher diversity of participants. For example, female participants, who are often underrepresented in VR research~\cite{Peck2020Mind}, were nearly equally represented in our sample.
Despite being shown only online videos, participants rated immersion as medium to high (see \autoref{sec:participants}). However, results with immersive MR headsets may differ from the results of our study, so future research should look into enhancing external validity. 
Further, research using actual MR setups in longitudinal studies is necessary to validate and extend the insights gained from this study. Such longitudinal studies also allow the inclusion of the high-level dark pattern of \textit{Obstruction} as it requires a longer user journey to manifest and, therefore, was not considered for the user study (see \autoref{sec:dark_patterns}). 
Further, we designed our dark pattern scenarios using the ontology by \citet{Gray.2024}, which was the most recent resource available at the time for designing these scenarios and conducting our user study. Because this taxonomy synthesizes several well-established classifications~\cite{Gray.2018, mathur_dark_2019, bosch_tales_2016, luguri_shining_2021}, we argue that it provides a holistic foundation and ensures consistency with previous research on deceptive interface design. However, \citet{krauss_what_2024} later extended this taxonomy for XR, proposing ten new dark patterns specific to immersive technologies. Although their work was published after our study design was finalized, we intuitively incorporated some of their ideas, such as \textit{Realism Surcharge}. In our user study, this was reflected through \textit{Forced Registration}, which required users to pay for additional services to restore a previously available experience. Nevertheless, \citet{krauss_what_2024} also identified additional XR-specific patterns we did not adopt, such as \textit{Spatial Imbalance of Options}, where the system strategically places interaction elements to favor or disfavor certain choices. Future research should, therefore, apply their XR-focused taxonomy to compare user impacts from these novel patterns against those in \citet{Gray.2024}, exploring whether newly discovered XR-specific manipulations have more severe effects on users than established dark patterns.

Additionally, we applied the four dark patterns to three augmentation targets: place, product, and person. Yet, dark patterns could be applied to a wider range of targets, such as buildings, points of interest, or stores. However, to reduce the complexity and length of the user study, we limited our investigation to these three augmentation targets. Similarly, we explored only one scenario per dark pattern. A single scenario per dark pattern limits our findings' generalizability, as the observed effects may partly result from the specific design rather than the inherent impact of the dark pattern. Thus, future research should test multiple designs (e.g., having different designers create their versions of each dark pattern). 
Moreover, all dark patterns applied to the person were framed in a romantic context, simulating a dating app~\cite{tinderverse}. We augmented a female person for this purpose. However, we did not collect information on participants' sexual orientation, which potentially made specific augmentations inappropriate for some participants. 
Further, participants' inexperience in MR applications might have biased their ratings. Yet, as MR is an emerging technology, many users will be inexperienced at first sight.

\section{Conclusion}
This paper explored four dark patterns—\textit{Emotional or Sensory Manipulation}, \textit{Forced Registration}, \textit{Hiding Information}, and \textit{Urgency}—within MR environments, applied to three different targets: a place, a product, and a person. We created 12 videos with dark patterns and one baseline video without dark patterns from the perspective of an individual wearing MR glasses during a city walk. Through a two-factorial online study involving N=74 participants, we assessed the impact of the dark pattern on potential users regarding comfort, intention to use, system darkness, and reactance.
While most previous work on dark patterns in MR used speculative co-design and qualitative expert workshops, we used quantitative measures to assess the impact of concrete dark patterns.
All dark patterns significantly reduced the user's comfort and intention to use MR while increasing reactance, with the most negative effects observed in scenarios involving personal interference (e.g., obscuring a person’s face) and monetary manipulation (e.g., sneaking products into a shopping cart). Additionally, we found that the dark patterns \textit{Emotional or Sensory Manipulation} and \textit{Hiding Information} yielded comparable impacts on users across all three augmentation targets. This similarity suggests that current classifications of dark patterns might need to consider their actual user impact alongside design techniques. Open feedback also revealed that participants experienced additional negative feelings, such as frustration and concerns about potential privacy risks. They also pointed out the potential dangers of obscuring real-world objects, which could lead to hazardous situations.
These insights highlight the importance of developing ethical and legal guidelines to detect and prevent dark patterns in MR, similar to what has been done for traditional interfaces, as this immersive technology continues to evolve.

\section*{Open Science}
Due to anonymity reasons, the full-length videos used in this study will be made available to interested researchers upon request. For a first look, please have a look at the paper's trailer video.


\begin{acks}
We would like to thank Sophia Ppali for her valuable feedback, insightful comments, and support in proofreading the manuscript. 

This work was partially funded by the Leibniz ScienceCampus Bremen Digital Public Health, which is jointly funded by the Leibniz Association (W72/2022), the Federal State of Bremen, and the Leibniz Institute for Prevention Research and Epidemiology – BIPS.
\end{acks}

\bibliographystyle{ACM-Reference-Format}
\bibliography{sample-base}

\onecolumn
\appendix

\section{Question Items Used in the User Study}\label{app:usedquestionaires}

\begin{table*}[ht!]
\fontsize{9pt}{8pt}\selectfont 
\caption{Question items used in the user study}
\begin{tabularx}{\textwidth}{p{1.4cm}p{8.2cm}p{2.5cm}X}

   \textbf{Measure.}& Question Item & Answer Items & Ref.   \\
   \addlinespace[2pt]
   \hline 
   \addlinespace[4pt]
   \textbf{Reactance}
   & 
   I want to be in control, not the augmentations \par
   \vspace{6.1pt}
   I like to act independently from the augmentations \par
   \vspace{6.1pt}
   I don't want the augmentations to tell me what to do \par
   \vspace{6.1pt}
   I don't let the augmentations impose their will on me \par
   \vspace{6.1pt}
   I alone determine what I do, not the augmentations \par
   \vspace{6.1pt}
   The augmentations frustrate me \par
   \vspace{6.1pt}
   The augmentations make me angry \par
   \vspace{6.1pt}
   I get mad when I have to interact with the augmentations \par
   \vspace{6.1pt}
   When I only see the augmentations, I burst with anger \par
   \vspace{6.1pt}
   The augmentations are simply bad! \par
   \vspace{6.1pt}
   At best, I would like to never see or hear anything about the augmentations again \par
   \vspace{6.1pt}
   I don't even want the augmentations as a present \par
   \vspace{6.1pt}
   The whole concept of the augmentations is a poor approach \par
   &    5-point Likert scale from (1)
 ``strongly disagree'', to (5) ``strongly agree'' & \cite{Ehrenbrink.2020} \\

   \hline 
   \addlinespace[4pt]
   \textbf{System Darkness Scale}
   & The augmentations tricked me into performing certain actions I did not intend to do \par
   \vspace{6.1pt}
   The augmentations performed certain actions I was not aware of \par
   \vspace{6.1pt}
   The augmentations pushed me into spending more money than I originally anticipated	\par
   \vspace{6.1pt}
   The augmentations performed certain actions without my consent	\par
   \vspace{6.1pt}
   I felt deceived/misled by the augmentations \par
   & 5-point Likert scale from (1)
 ``strongly disagree'', to (5) ``strongly agree'' & \cite{van_nimwegen_shedding_2022} \\

   \hline 
   \addlinespace[5pt]
   \textbf{Comfort}
   & Please rate how comfortable you felt with the augmentations \newline
   & 5-point Likert scale from (1) ``very uncomfortable'' to (5) ``very comfortable'' \par & \cite{rixen_exploring_2021} \\  
   
   \hline 
   \addlinespace[4pt]
   \textbf{Intention to Use}
   & Assuming I have access to the system, I intend to use it \par
   \vspace{6.1pt}
   Given that I have access to the system, I predict that I would use it \par
   & 7-point Likert scale from (1), ``Unlikely'', to (7), ``Likely'' & \cite{venkatesh_theoretical_2000} \\

   \hline 
   \addlinespace[4pt]
   \textbf{Immersion}
   & In the virtual simulation, I could forget my real problems. \par
   \vspace{6.1pt}
   When I use the virtual simulation, I feel like I'm in another world. \par
   \vspace{6.1pt}
   During the virtual simulation, I completely forgot the world around me. \par
   \vspace{6.1pt}
   Through the virtual simulation, I had the feeling of really experiencing the situation.
   & 7-point Likert scale from (1), ``Totally Disagree'', to (7), ``Totally Agree'' & \cite{kothgassner2013technology} \\

\label{tab:question_items}
\end{tabularx}

\end{table*}

\onecolumn
\newpage

\section{Descriptive Data of the User Study}\label{app:descriptive}
\begin{table*}[ht]
\caption{Table of the descriptive data of the user study}
\begingroup 
\fontsize{8pt}{9pt}\selectfont 
\begin{tabularx}{\textwidth}{lllrrrr}
 \textbf{Variable} & \textbf{Dark Pattern} & \textbf{Augmentation} & \textbf{min} & \textbf{max} & \textbf{mean} & \textbf{sd} \\ 
  \hline
  \addlinespace[3pt]
        Reactance~\cite{Ehrenbrink.2020} & Emotional or Sensory Manipulation & Place & 1.62 & 5.00 & 3.42 & 1.00 \\
                  & Emotional Manipulation or Sensory & Product & 1.00 & 5.00 & 3.60 & 1.09 \\
                  & Emotional Manipulation or Sensory & Person & 1.00 & 5.00 & 3.24 & 1.06 \\ \cline{2-7}
                  & Forced Registration & Place & 1.00 & 5.00 & 3.46 & 1.15 \\
                  & Forced Registration & Product & 1.00 & 5.00 & 3.63 & 1.09 \\
                  & Forced Registration & Person & 1.00 & 5.00 & 3.72 & 1.06 \\ \cline{2-7}
                  & Hiding Information & Place & 1.00 & 5.00 & 3.54 & 1.12 \\
                  & Hiding Information & Product & 1.46 & 5.00 & 3.85 & 1.01 \\
                  & Hiding Information & Person & 1.54 & 5.00 & 3.40 & 1.07 \\ \cline{2-7}
                  & Urgency & Place & 1.00 & 5.00 & 3.49 & 1.05 \\
                  & Urgency & Product & 1.00 & 5.00 & 3.44 & 1.09 \\
                  & Urgency & Person & 1.00 & 5.00 & 3.40 & 1.08 \\ \cline{2-7}
                  & Baseline &  & 1.00 & 5.00 & 2.77 & 0.85 \\ 
                  \hline
                    \addlinespace[3pt]
        System Darkness~\cite{van_nimwegen_shedding_2022} & Emotional or Sensory Manipulation & Place & 0 & 100 & 41.10 & 26.30 \\
                  & Emotional or Sensory Manipulation& Product & 0 & 100 & 48.60 & 26.80 \\
                  & Emotional or Sensory Manipulation& Person & 0 & 100 & 39.90 & 25.60 \\ \cline{2-7}
                  & Forced Registration & Place & 0 & 100 & 43.20 & 30.90 \\
                  & Forced Registration & Product & 0 & 100 & 47.00 & 31.60 \\
                  & Forced Registration & Person & 0 & 100 & 45.70 & 30.30 \\ \cline{2-7}
                  & Hiding Information & Place & 0 & 100 & 55.00 & 31.00 \\
                  & Hiding Information & Product & 0 & 100 & 70.00 & 31.30 \\
                  & Hiding Information & Person & 0 & 100 & 47.40 & 27.30 \\ \cline{2-7}
                  & Urgency & Place & 0 & 100 & 46.20 & 27.10 \\
                  & Urgency & Product & 0 & 100 & 48.00 & 27.90 \\
                  & Urgency & Person & 0 & 100 & 41.10 & 25.90 \\ \cline{2-7}
                  & Baseline &  & 0 & 80 & 21.76 & 21.64 \\ 
                  \hline
                    \addlinespace[3pt]
        Intention to Use~\cite{venkatesh_theoretical_2000} & Emotional or Sensory Manipulation & Place & 1.00 & 7.00 & 3.75 & 2.21 \\
                  & Emotional or Sensory Manipulation & Product & 1.00 & 7.00 & 3.41 & 2.22 \\
                  & Emotional or Sensory Manipulation & Person & 1.00 & 7.00 & 3.77 & 2.31 \\ \cline{2-7}
                  & Forced Registration & Place & 1.00 & 7.00 & 3.43 & 2.25 \\
                  & Forced Registration & Product & 1.00 & 7.00 & 3.05 & 2.29 \\
                  & Forced Registration & Person & 1.00 & 7.00 & 2.94 & 2.29 \\ \cline{2-7}
                  & Hiding Information & Place & 1.00 & 7.00 & 3.42 & 2.20 \\
                  & Hiding Information & Product & 1.00 & 7.00 & 2.96 & 2.25 \\
                  & Hiding Information & Person & 1.00 & 7.00 & 3.41 & 2.18 \\ \cline{2-7}
                  & Urgency & Place & 1.00 & 7.00 & 3.48 & 2.13 \\
                  & Urgency & Product & 1.00 & 7.00 & 3.40 & 2.26 \\
                  & Urgency & Person & 1.00 & 7.00 & 3.39 & 2.26 \\ \cline{2-7}
                  & Baseline &  & 1.00 & 7.00 & 4.67 & 2.01 \\ 
                  \hline
                    \addlinespace[3pt]
        Comfort~\cite{rixen_exploring_2021} & Emotional or Sensory Manipulation & Place & 1.00 & 5.00 & 2.96 & 1.19 \\
                  & Emotional or Sensory Manipulation & Product & 1.00 & 5.00 & 2.59 & 1.18 \\
                  & Emotional or Sensory Manipulation & Person & 1.00 & 5.00 & 2.74 & 1.35 \\ \cline{2-7}
                  & Forced Registration & Place & 1.00 & 5.00 & 2.61 & 1.32 \\
                  & Forced Registration & Product & 1.00 & 5.00 & 2.43 & 1.23 \\
                  & Forced Registration & Person & 1.00 & 5.00 & 2.18 & 1.21 \\ \cline{2-7}
                  & Hiding Information & Place & 1.00 & 5.00 & 2.43 & 1.33 \\
                  & Hiding Information & Product & 1.00 & 5.00 & 2.16 & 1.29 \\
                  & Hiding Information & Person & 1.00 & 5.00 & 2.54 & 1.31 \\ \cline{2-7}
                  & Urgency & Place & 1.00 & 5.00 & 2.66 & 1.21 \\
                  & Urgency & Product & 1.00 & 5.00 & 2.74 & 1.21 \\
                  & Urgency & Person & 1.00 & 5.00 & 2.47 & 1.27 \\ \cline{2-7}
                  & Baseline &  & 1.00 & 5.00 & 3.59 & 1.01 \\ 
                  \hline
                    \addlinespace[3pt]

\label{tab:_descr_stat}
\end{tabularx}
\endgroup 
\end{table*}

\end{document}